\documentclass[12pt]{article}
\usepackage{amssymb}

\usepackage{graphicx}
\usepackage{amsmath}
\usepackage{makeidx}
\usepackage{indentfirst}

\newcounter{resultnum}[section]

\newcounter{conclusionnum}[section]

\newcounter{conditionnum}[section]

\newcounter{conjecturenum}[section]

\newcounter{examplenum}[section]

\newcounter{exercisenum}[section]

\newcounter{lemmanum}[section]

\newcounter{notationnum}[section]

\newcounter{theoremnum}[section]

\newcounter{definitionnum}[section]

\newcounter{corollarynum}[section]

\newcounter{remarknum}[section]

\newcounter{propositionnum}[section]

\newcounter{acknowledgementnum}[section]

\newcounter{algorithmnum}[section]

\newcounter{axiomnum}[section]

\newcounter{casenum}[section]

\newcounter{claimnum}[section]

\newcounter{summarynum}[section]

\newcounter{problemnum}[section]

\begin{document}

\title{Constant Curvature Coefficients and \\
Exact Solutions in Fractional Gravity\\
and Geometric Mechanics}
\date{March 30, 2011}
\author{\textbf{Dumitru Baleanu}\thanks{%
On leave of absence from Institute of Space Sciences, P. O. Box, MG-23, R
76900, Magurele--Bucharest, Romania;\ dumitru@cancaya.edu.tr,
baleanu@venus.nipne.ro} \\
\textsl{\small Department of Mathematics and Computer Sciences,} \\
\textsl{\small \c Cankaya University, 06530, Ankara, Turkey } \\
\and 
\textbf{Sergiu I. Vacaru} \thanks{%
sergiu.vacaru@uaic.ro;\ http://www.scribd.com/people/view/1455460-sergiu}
\and \textsl{\small Science Department, University "Al. I. Cuza" Ia\c si, }
\\
\textsl{\small 54, Lascar Catargi street, Ia\c si, Romania, 700107}}
\maketitle

\begin{abstract}
We study fractional
configurations in gravity theories and Lagrange mechanics. The approach is based on Caputo fractional derivative which gives zero  for actions on constants. We elaborate
fractional geometric models of physical interactions and we formulate a method of nonholonomic deformations to other types of fractional derivatives. The  main result of this paper consists in a proof that for corresponding classes of  nonholonomic distributions a large class of physical theories are modelled as nonholonomic manifolds with constant matrix curvature. This allows us to encode the fractional dynamics of interactions and constraints into the geometry of curve flows and solitonic hierarchies.

\vskip5pt

\textbf{Keywords:}\ fractional calculus, fractional geometry, fractional
gravity, fractional Lagrange mechanics, generalized Finsler geometry,
nonlinear connection, nonholonomic manifold.

\vskip3pt

 PACS2010:\ 02.30.Ik, 45.10Hj, 45.10.Na, 05.45.Yv, 02.40.Yy,
45.20.Jj, 04.20.Jb, 04.50.Kd
\end{abstract}


\section{Introduction}

It is well known that both the modified Korteweg -- de Vries, mKdV, and the
sine--Gordon, SG, (solitonic) equations can be encoded as flows of the
curvatures invariant of plane curves in Euclidean plane geometry \cite%
{lamb,goldstein,nakayama,langer,chou,befa}. Such constructions were
developed \cite{athorne,saw,anc2} for curve flows in Riemannian manifolds of
constant curvature \cite{helgason,sharpe} which gave rise, for instance, to
a vector generalization of the mKdV equation and encode its bi--Hamiltonian
structure of multi--component mKdV equations.

A crucial condition behind above mentioned geometric models
was that the frame curvature matrix is constant for a series of examples of
Riemann symmetric spaces, with group symmetries etc, for which certain
classes of nonlinear/ solitonic equations can be associated. It was
considered that the approach would have a limited importance, for instance,
for study of more general classes of curved spaces (pseudo--Riemanian ones,
Einstein manifolds etc) if such geometries are not with constant curvature
matrix coefficients. Nevertheless, it was possible to elaborate a similar
method of solitonic encoding of geometric data for generalized geometries
with non--constant curvatures using certain ideas and methods from the
geometry of nonholonomic distributions and modelling Lagrange--Finsler
geometries \cite{vrflg}. Such models with nontrivial nonlinear connection
structure were elaborated on (co) tangent bundles (see \cite{ma1} and
references therein) and our proposal was to apply such methods on (pseudo)
Riemannian and Einstein manifolds endowed with nonholonomic distributions.

We developed a geometric formalism when nonholonomic deformations of
geometric structures determined by a fundamental Lagrange/ Finsler /
Hamilton generating function (or, for instance, and Einstein metric) induce
a canonical connection, adapted to a necessary type nonlinear connection
structure, for which the matrix coefficients of curvature are constant \cite%
{vacap,vanco}. For such an auxiliary connection, it is possible to define a
bi--Hamiltonian structure and derive the corresponding solitonic hierarchy.

As it is known the fractional calculus which deals with derivative and integrals of arbitrary orders \cite{trujillo,samko,podlubny1,richard,13}  is an emerging field and it has many applications in various fields of science and engineering \cite{agrawal,klimek,baleanu1,baleanu2,13a,13b,13,blochmb,machado3}.

In a series of our works \cite{vrfrf,vrfrg,bv1,bv2} (we recommend the reader
to see details, discussions and bibliography, including notation conventions
etc), we proposed program of geometrization of fractional calculus applications in classical and quantum fractional mechanics, field
theories and Ricci flow evolution theories with fractional/noncommutative/
stochastic etc derivatives. It should be emphasized here that it is not
possible to elaborate an unified geometric formalism for all types of
existing definitions of fractional derivatives. Different such operators
have their priorities and minuses and result in very different geometric
(integro--differential) and, for instance, physical implications. Nevertheless, there is a class of fractional derivatives resulting in zero
for actions on constants. This property is crucial for constructing
geometric models of physical and mechanical theories with fractional
calculus which (from many formal points of view) are very similar to those
on integer dimensional spaces but with corresponding nonholonomic
distributions selecting a prescribed type of fractional dynamics and/or
evolution and, in general, nonholonomic constraints. Such an analogy allows
us to involve in our study a number of very powerful geometric methods
already used, for instance, in deformation quantization, geometric
mechanics, Lagrange--Finsler geometry etc. Via nonholonomic
deformations/transforms, the constructions based on Caputo fractional
calculus seem to have generalizations to include other types of fractional
derivatives.

The main goal of this paper is to provide such a geometrization of
fractional calculus formalism
\cite{trujillo,samko,podlubny1,richard,13,13a,13b,mainardi3,agrawal,baleanu1,baleanu2}
extended to certain applications/modifications of gravity and
Lagrange mechanics theories, in a fractional calculus  manner,
which will allow in the second our partner work \cite{bv4} to
encode such geometric constructions into (fractional)
bi--Hamiltonians and
associated solitonic hierarchies. In brief, two of our works (this one and \cite%
{bv4}) present a fractional version of articles
\cite{vacap,vanco}.

The paper is organized as follows:

In section 2, we outline the geometry of N--adapted fractional spaces. In
section 3, we provide an introduction into the theory of fractional gravity,
show how the fundamental equations in such theories can be integrated in
general form and provide a geometrization for fractional Lagrange mechanics.
Section 4 is devoted to the geometry of fractional curve flows and related
models of fractional N--anholonomic Klein spaces. The Appendix contains
necessary definitions and formulas on Caputo fractional derivatives and
related nonholonomic fractional differential geometry (with nonlinear
connections).

\section{Fractional Manifolds with Constant Curvature Coefficients}

Let us consider a ''prime'' nonholonomic manifold $\mathbf{V}$ is of integer
dimension $\dim $ $\mathbf{V}=n+m,n\geq 2,m\geq 1.$\footnote{%
A nonholonomic manifold is a manifold endowed with a non--integrable
(equivalently, nonholonomic, or anholonomic) distribution. There are three
useful (for our considerations) examples when 1) $\mathbf{V}$ is a (pseudo)
Riemannian manifold; 2) $\mathbf{V}=E(M),$ or 3)\ $\mathbf{V}=TM,$ for a
vector, or tangent, bundle on a base manifold $M.$ We also emphasize that in
this paper we follow the conventions from Refs. \cite{vrflg,vrfrf,vrfrg}
when left indices are used as labels and right indices may be abstract ones
or running certain values.}\ Its fractional extension $\overset{\alpha }{%
\mathbf{V}}$ is modelled by a quadruple $(\mathbf{V},\overset{\alpha }{%
\mathbf{N}},\overset{\alpha }{\mathbf{d}},\overset{\alpha }{\mathbf{I}}),$
where $\overset{\alpha }{\mathbf{N}}$  is a nonholonomic distribution
stating a nonlinear connection (N--connection) structure (for details, see
Appendix \ref{sappa} with explanations for formula (\ref{whit}). The
fractional differential structure $\overset{\alpha }{\mathbf{d}}$ is
determined by Caputo fractional derivative (\ref{lfcd}) following formulas (%
\ref{frlcb}) and (\ref{frlccb}).
The non--integer integral structure $%
\overset{\alpha }{\mathbf{I}}$ is defined by rules of type (\ref{aux01}).

The goal of this section is to prove that it is possible to construct a
metric compatible linear connection $\widetilde{\mathbf{D}}$ with constant
matrix coefficients of curvature, computed with respect to 'N--adapted'
frames, for any fractional metric$\ \overset{\alpha }{\mathbf{g}}$ (\ref%
{fmcf}) (equivalently, (\ref{m1})) on a nonholonomic manifold $\overset{%
\alpha }{\mathbf{V}}.$

\subsection{N--adapted frame transforms and fractional metrics}

For any respective frame and co--frame (dual) structures,\newline
$\ ^{\alpha }e_{\alpha ^{\prime }}=(\ ^{\alpha }e_{i^{\prime }},\ ^{\alpha
}e_{a^{\prime }})$ and $\ ^{\alpha }e_{\ }^{\beta ^{\prime }}=(\ ^{\alpha
}e^{i^{\prime }},\ ^{\alpha }e^{a^{\prime }})$ \ on $\overset{\alpha }{%
\mathbf{V}}\mathbf{,}$ we can consider \ frame transforms
\begin{equation}
\ ^{\alpha }e_{\alpha }=A_{\alpha }^{\ \alpha ^{\prime }}(x,y)\ ^{\alpha
}e_{\alpha ^{\prime }}\mbox{\ and\ }\ ^{\alpha }e_{\ }^{\beta }=A_{\ \beta
^{\prime }}^{\beta }(x,y)\ ^{\alpha }e^{\beta ^{\prime }}.  \label{nhft}
\end{equation}

A subclass of frame transforms (\ref{nhft}), for fixed ''prime'' and
''target'' frame structures, is called N--adapted if such nonholonomic
transformations preserve the splitting defined by a N--connection structure $%
\mathbf{N}=\{N_{i}^{a}\}.$

Under (in general, nonholonomic) frame transforms, the metric coefficients
of any metric structure $\overset{\alpha }{\mathbf{g}}$ on $\overset{\alpha }%
{\mathbf{V}}$ are re--computed following formulas
\begin{equation*}
\ ^{\alpha }g_{\alpha \beta }(x,y)=A_{\alpha }^{~\ \alpha ^{\prime
}}(x,y)~A_{\beta }^{\ \beta ^{\prime }}(x,y)\ ^{\alpha }g_{\alpha ^{\prime
}\beta ^{\prime }}(x,y).
\end{equation*}

For any fixed \ $\overset{\alpha }{\mathbf{g}}$ and $\overset{\alpha }{%
\mathbf{N}},$ there are N--adapted frame transforms when
\begin{eqnarray*}
\overset{\alpha }{\mathbf{g}} &=&\ ^{\alpha }g_{ij}(x,y)\ \ ^{\alpha
}e^{i}\otimes \ ^{\alpha }e^{j}+\ ^{\alpha }h_{ab}(x,y)\ ^{\alpha }\mathbf{e}%
^{a}\otimes \ ^{\alpha }\mathbf{e}^{b}, \\
&=&\ ^{\alpha }g_{i^{\prime }j^{\prime }}(x,y)\ \ ^{\alpha }e^{i^{\prime
}}\otimes \ ^{\alpha }e^{j^{\prime }}+\ ^{\alpha }h_{a^{\prime }b^{\prime
}}(x,y)\ \ ^{\alpha }\mathbf{e}^{a^{\prime }}\otimes \ ^{\alpha }\mathbf{e}%
^{b^{\prime }},
\end{eqnarray*}%
where $\ ^{\alpha }\mathbf{e}^{a}$ and $\ ^{\alpha }\mathbf{e}^{a^{\prime }}$
are elongated following formulas (\ref{ddif}), respectively by $\ ^{\alpha
}N_{\ j}^{a}$ and
\begin{equation}
\ ^{\alpha }N_{\ j^{\prime }}^{a^{\prime }}=A_{a}^{~\ a^{\prime }}(x,y)A_{\
j^{\prime }}^{j}(x,y)\ ^{\alpha }N_{\ j}^{a}(x,y),  \label{ncontri}
\end{equation}%
or, inversely,
\begin{equation*}
\ ^{\alpha }N_{\ j}^{a}=A_{a^{\prime }}^{~\ a}(x,y)A_{\ j}^{j^{\prime
}}(x,y)\ ^{\alpha }N_{\ j^{\prime }}^{a^{\prime }}(x,y)
\end{equation*}%
with prescribed $\ ^{\alpha }N_{\ j^{\prime }}^{a^{\prime }}.$

We preserve the N--connection splitting for any frame transform of type (\ref%
{nhft}) when
\begin{equation*}
\ ^{\alpha }g_{i^{\prime }j^{\prime }}=A_{\ i^{\prime }}^{i}A_{\ j^{\prime
}}^{j}\ ^{\alpha }g_{ij},\ \ ^{\alpha }h_{a^{\prime }b^{\prime }}=A_{\
a^{\prime }}^{a}A_{\ b^{\prime }}^{b}\ ^{\alpha }h_{ab},
\end{equation*}%
for $A_{i}^{~\ i^{\prime }}$ constrained to get holonomic $\ ^{\alpha
}e^{i^{\prime }}=A_{i}^{~\ i^{\prime }}\ ^{\alpha }e^{i},$ i.e. $[\ ^{\alpha
}e^{i^{\prime }},\ ^{\alpha }e^{j^{\prime }}]=0$ and $\ ^{\alpha }\mathbf{e}%
^{a^{\prime }}=dy^{a^{\prime }}+\ ^{\alpha }N_{\ j^{\prime }}^{a^{\prime
}}dx^{j^{\prime }},$ for certain $x^{i^{\prime }}=x^{i^{\prime
}}(x^{i},y^{a})$ and $y^{a^{\prime }}=y^{a^{\prime }}(x^{i},y^{a}),$ with $\
^{\alpha }N_{\ j^{\prime }}^{a^{\prime }}$ computed following formulas (\ref%
{ncontri}). Such conditions can be satisfied by prescribing from the very
beginning a nonholonomic distribution of necessary type. The constructions
can be equivalently inverted, when $\ ^{\alpha }g_{\alpha \beta }$ and $\
^{\alpha }N_{i}^{a}$ are computed from $\ ^{\alpha }g_{\alpha ^{\prime
}\beta ^{\prime }}$ and $\ ^{\alpha }N_{i^{\prime }}^{a^{\prime }},$ if both
the metric and N--connection splitting structures are fixed on $\overset{%
\alpha }{\mathbf{V}}.$

\subsection{d--connections with constant curvature coefficients}

From the class of metric compatible fractional d--connections\footnote{%
see definitions and main formulas in Appendix \ref{sappa}} uniquely defined
by a fractional metric structure $\overset{\alpha }{\mathbf{g}},$ we chose
such a N--connection splitting with nontrivial coefficients $\ ^{\alpha
}N_{i}^{a}(x,y)$ when with respect to a N--adapted frame the canonical
d--connection (\ref{candcon}) has constant coefficients.

Any fractional metric $\overset{\alpha }{\mathbf{g}}$ on $\overset{\alpha }{%
\mathbf{V}}$ defines a set of metric compatible fractional d--connections of
type
\begin{equation}
\ _{0}^{\alpha }\widetilde{\mathbf{\Gamma }}_{\ \alpha ^{\prime }\beta
^{\prime }}^{\gamma ^{\prime }}=\left( \ ^{\alpha }\widehat{L}_{j^{\prime
}k^{\prime }}^{i^{\prime }}=0,\ ^{\alpha }\widehat{L}_{b^{\prime }k^{\prime
}}^{a^{\prime }}=\ _{0}^{\alpha }\widehat{L}_{b^{\prime }k^{\prime
}}^{a^{\prime }}=const,\ ^{\alpha }\widehat{C}_{j^{\prime }c^{\prime
}}^{i^{\prime }}=0,\ ^{\alpha }\widehat{C}_{b^{\prime }c^{\prime
}}^{a^{\prime }}=0\right)  \label{ccandcon}
\end{equation}%
with respect to correspondingly constructed N--adapted frames (\ref{dder})
and (\ref{ddif}), when $\ \ ^{\alpha }\mathbf{N}=\{\ ^{\alpha }N_{i^{\prime
}}^{a^{\prime }}(x,y)\}$ is a nontrivial solution of the system of equations%
\begin{equation}
2\ _{0}^{\alpha }\widehat{L}_{b^{\prime }k^{\prime }}^{a^{\prime }}=\overset{%
\alpha }{\underline{\partial }}_{b^{\prime }}(\ ^{\alpha }N_{k^{\prime
}}^{a^{\prime }})-\ _{0}^{\ \alpha }h^{a^{\prime }c^{\prime }}\ _{0}^{\
\alpha }h_{d^{\prime }b^{\prime }}\ \overset{\alpha }{\underline{\partial }}%
_{c^{\prime }}\ ^{\alpha }N_{k^{\prime }}^{d^{\prime }}  \label{auxf1}
\end{equation}%
for any nondegenerate constant--coefficients symmetric matrix $\ _{0}^{\
\alpha }h_{d^{\prime }b^{\prime }}$ and its inverse $\ \ _{0}^{\ \alpha
}h^{a^{\prime }c^{\prime }}.$ \ The Caputo fractional derivative $\overset{%
\alpha }{\underline{\partial }}_{b^{\prime }}$ allows us to dub such
constructions for fractional spaces \cite{vacap,vanco}.

For both integer and non--integer dimensions, the coefficients $\
_{\shortmid }^{\alpha }\Gamma _{\ \alpha ^{\prime }\beta ^{\prime }}^{\gamma
^{\prime }}$ of the corresponding to $\overset{\alpha }{\mathbf{g}}$
Levi--Civita connection $\ _{\mathbf{g}}^{\alpha }\nabla $ are not constant
with respect to N--adapted frames.\footnote{%
In explicit form, such coefficients are computed following formulas (\ref%
{cdeft}).}

The curvature tensor of fractional d--connection $\ _{0}^{\alpha }\widetilde{%
\mathbf{\Gamma }}_{\ \alpha ^{\prime }\beta ^{\prime }}^{\gamma ^{\prime }}$
(\ref{ccandcon}) defined by a metric $\overset{\alpha }{\mathbf{g}}$ has
constant coefficients with respect to N--adapted frames $\ ^{\alpha }\mathbf{%
e}_{\alpha ^{\prime }}=[\ ^{\alpha }\mathbf{e}_{i^{\prime }},\ ^{\alpha
}e_{a^{\prime }}]$ and $\ ^{\alpha }\mathbf{e}^{\alpha ^{\prime }}=[\
^{\alpha }e^{i^{\prime }},\ ^{\alpha }\mathbf{e}^{a^{\prime }}]$ with $\
^{\alpha }N_{k^{\prime }}^{d^{\prime }}$ subjected to conditions (\ref{auxf1}%
). \ Introducing constant coefficients $\ _{0}^{\alpha }\widetilde{\mathbf{%
\Gamma }}_{\ \alpha ^{\prime }\beta ^{\prime }}^{\gamma ^{\prime }}$ (\ref%
{ccandcon}) into formulas (\ref{tors}), we get
\begin{eqnarray*}
&&\ _{0}^{\alpha }\widetilde{\mathbf{R}}_{\ \beta ^{\prime }\gamma ^{\prime
}\delta ^{\prime }}^{\alpha ^{\prime }}=(\ _{0}^{\alpha }\widetilde{R}%
_{~h^{\prime }j^{\prime }k^{\prime }}^{i^{\prime }}=0,\ _{0}^{\alpha }%
\widetilde{R}_{~b^{\prime }j^{\prime }k^{\prime }}^{a^{\prime }}=\
_{0}^{\alpha }\widehat{L}_{\ b^{\prime }j^{\prime }}^{c^{\prime }}\
_{0}^{\alpha }\widehat{L}_{\ c^{\prime }k^{\prime }}^{a^{\prime }}-\
_{0}^{\alpha }\widehat{L}_{\ b^{\prime }k^{\prime }}^{c^{\prime }}\
_{0}^{\alpha }\widehat{L}_{\ c^{\prime }j^{\prime }}^{a^{\prime }} \\
&&= const,\ _{0}^{\alpha }\widetilde{P}_{~h^{\prime }j^{\prime }a^{\prime
}}^{i^{\prime }}=0,\ _{0}^{\alpha }\widetilde{P}_{~b^{\prime }j^{\prime
}a^{\prime }}^{c^{\prime }}=0,\ _{0}^{\alpha }\widetilde{S}_{~j^{\prime
}b^{\prime }c^{\prime }}^{i^{\prime }}=0,\ _{0}^{\alpha }\widetilde{S}%
_{~b^{\prime }d^{\prime }c^{\prime }}^{a^{\prime }}=0).
\end{eqnarray*}%
Such proprieties hold for a class of prescribed nonholonomic distributions.
Of course, in general, with respect to local coordinate (or other
N--adapted) frames, the curvature d--tensor $\ \ ^{\alpha }\widehat{\mathbf{R%
}}_{\ \beta \gamma \delta }^{\alpha }$ does not have constant coefficients.
Nevertheless, we can always chose certain nonholonomic N--connection
configurations when the corresponding adapting will generate d--connections
with constant matrix curvature coefficients.\footnote{%
Using deformation relation (\ref{cdeft}), we can compute the corresponding
Ricci tensor $\ _{\shortmid }^{\alpha }R_{\ \beta \gamma \delta }^{\alpha }$
for the Levi--Civita connection $\ \ _{\mathbf{g}}^{\alpha }\nabla ,$ which
is a general one with 'non-constant' coefficients with respect to any local
frames.}

Using formulas (\ref{dricci}) and (\ref{sdccurv}), we compute
\begin{eqnarray*}
\ _{0}^{\alpha }\overleftrightarrow{\mathbf{R}} &\doteqdot &\ _{0}^{\alpha }%
\mathbf{g}^{\alpha ^{\prime }\beta ^{\prime }}\ _{0}^{\alpha }\widetilde{%
\mathbf{R}}_{\alpha ^{\prime }\beta ^{\prime }}=\ _{0}^{\alpha }g^{i^{\prime
}j^{\prime }}\ _{0}^{\alpha }\widetilde{R}_{i^{\prime }j^{\prime }}+\
_{0}^{\alpha }h^{a^{\prime }b^{\prime }}\ _{0}^{\alpha }\widetilde{S}%
_{a^{\prime }b^{\prime }} \\
&=&\ _{0}^{\alpha }\overrightarrow{R}+\ _{0}^{\alpha }\overleftarrow{S}%
=const.
\end{eqnarray*}
We conclude that a fractional d--connection $\ _{0}^{\alpha }\widetilde{%
\mathbf{\Gamma }}_{\ \alpha ^{\prime }\beta ^{\prime }}^{\gamma ^{\prime }}$
(\ref{ccandcon}) has constant scalar curvature.

\section{Fractional Einstein and/or Lagrange Spaces}

The goal of this section is to show how the integer dimensional Einstein
gravity and Lagrange mechanics can be encoded and/or generalized in terms of
fundamental geometric objects on nonholonomic fractional manifolds. We
review and develop our constructions from \cite{vrfrf,vrfrg,bv1,bv2}.

\subsection{Fractional gravity}

An unified approach to Einstein--Lagrange/Finsler gravity for arbitrary
integer and non--integer dimensions is possible for the fractional canonical
d--connection $\ ^{\alpha }\widehat{\mathbf{D}}.$ The fractional
gravitational field equations are formulated for the Einstein d--tensor (\ref%
{enstdt}), following the same principle of constructing the matter source $\
^{\alpha }\mathbf{\Upsilon }_{\beta \delta }$ as in general relativity but
for fractional metrics and d--connections,%
\begin{equation}
\ ^{\alpha }\widehat{\mathbf{E}}_{\ \beta \delta }=\ ^{\alpha }\mathbf{%
\Upsilon }_{\beta \delta }.  \label{fdeq}
\end{equation}%
Such a system of integro--differential equations for generalized connections
can be restricted to fractional nonholonomic configurations for $\ ^{\alpha
}\nabla $ if we impose the additional constraints%
\begin{equation}
\ ^{\alpha }\widehat{L}_{aj}^{c}=\ ^{\alpha }e_{a}(\ ^{\alpha }N_{j}^{c}),\
\ ^{\alpha }\widehat{C}_{jb}^{i}=0,\ \ ^{\alpha }\Omega _{\ ji}^{a}=0.
\label{frconstr}
\end{equation}

There are not theoretical or experimental evidences that for fractional
dimensions we must impose conditions of type (\ref{frconstr}) but they have
certain physical motivation if we develop models which in integer limits
result in the general relativity theory.

\subsubsection{Separation of equations for fractional and integer dimensions}

We studied in detail \cite{vrfrg} what type of conditions must
satisfy the coefficients of a metric (\ref{m1}) for generating
exact solutions of the fractional Einstein equations (\ref{fdeq}).
For simplicity, we can use a ''prime'' dimension splitting of type
$2+2$ when coordinated are labelled in the form $u^{\beta
}=(x^{j},y^{3}=v,y^{4}),$ for $i,j,...=1,2.$ and the metric ansatz
has one Killing symmetry when the coefficients do not depend
explicitly on variable $y^{4}.$

 The solutions of equations can be constructed for a general source of type%
\footnote{%
such parametrizations of energy--momentum tensors are quite general ones for
\ various types of matter sources}
\begin{equation*}
\ ^{\alpha }\Upsilon _{\ \ \beta }^{\alpha }=diag[\mathbf{\ }%
^{\alpha }\Upsilon _{\gamma };\mathbf{\ }^{\alpha }\Upsilon _{1}=\mathbf{\ }%
^{\alpha }\Upsilon _{2}=\mathbf{\ }^{\alpha }\Upsilon _{2}(x^{k},v);\mathbf{%
\ }^{\alpha }\Upsilon _{3}=\mathbf{\ }^{\alpha }\Upsilon _{4}=\mathbf{\ }%
^{\alpha }\Upsilon _{4}(x^{k})]  \label{source}
\end{equation*}%
For such sources and ansatz with Killing symmetries for metrics, the
Einstein equations (\ref{fdeq}) transform into a system of partial
differential equations with separation of equations which can be integrated
in general form,
\begin{eqnarray}
\mathbf{\ }^{\alpha }\widehat{R}_{1}^{1} &=&\mathbf{\ }^{\alpha }\widehat{R}%
_{2}^{2}=-\frac{1}{2\mathbf{\ }^{\alpha }g_{1}\mathbf{\ }^{\alpha }g_{2}}%
\times \lbrack \mathbf{\ }^{\alpha }g_{2}^{\bullet \bullet }-\frac{\mathbf{\
}^{\alpha }g_{1}^{\bullet }\mathbf{\ }^{\alpha }g_{2}^{\bullet }}{2\mathbf{\
}^{\alpha }g_{1}}  \label{eq1} \\
&&-\frac{\left( \mathbf{\ }^{\alpha }g_{2}^{\bullet }\right) ^{2}}{2\mathbf{%
\ }^{\alpha }g_{2}}+\mathbf{\ }^{\alpha }g_{1}^{\prime \prime }-\frac{%
\mathbf{\ }^{\alpha }g_{1}^{\prime }\mathbf{\ }^{\alpha }g_{2}^{\prime }}{2%
\mathbf{\ }^{\alpha }g_{2}}-\frac{\left( \mathbf{\ }^{\alpha
}g_{1}^{^{\prime }}\right) ^{2}}{2\mathbf{\ }^{\alpha }g_{1}}]=-\mathbf{\ }%
^{\alpha }\Upsilon _{4},  \notag \\
\mathbf{\ }^{\alpha }\widehat{R}_{3}^{3} &=&\mathbf{\ }^{\alpha }\widehat{R}%
_{4}^{4}=-\frac{1}{2\mathbf{\ }^{\alpha }h_{3}\mathbf{\ }^{\alpha }h_{4}}[%
\mathbf{\ }^{\alpha }h_{4}^{\ast \ast }  \label{eq2} \\
&&-\frac{\left( \mathbf{\ }^{\alpha }h_{4}^{\ast }\right) ^{2}}{2\mathbf{\ }%
^{\alpha }h_{4}}-\frac{\mathbf{\ }^{\alpha }h_{3}^{\ast }\mathbf{\ }^{\alpha
}h_{4}^{\ast }}{2\mathbf{\ }^{\alpha }h_{3}}]=-\mathbf{\ }^{\alpha }\Upsilon
_{2},  \notag
\end{eqnarray}
\begin{eqnarray}
\mathbf{\ }^{\alpha }\widehat{R}_{3k} &=&\frac{\mathbf{\ }^{\alpha }w_{k}}{2%
\mathbf{\ }^{\alpha }h_{4}}\left[ \mathbf{\ }^{\alpha }h_{4}^{\ast \ast }-%
\frac{\left( \mathbf{\ }^{\alpha }h_{4}^{\ast }\right) ^{2}}{2\mathbf{\ }%
^{\alpha }h_{4}}-\frac{\mathbf{\ }^{\alpha }h_{3}^{\ast }\mathbf{\ }^{\alpha
}h_{4}^{\ast }}{2\mathbf{\ }^{\alpha }h_{3}}\right]  \label{eq3} \\
&&+\frac{\mathbf{\ }^{\alpha }h_{4}^{\ast }}{4\mathbf{\ }^{\alpha }h_{4}}%
\left( \frac{_{\ _{1}x^{i}}\overset{\alpha }{\underline{\partial }}_{x^{i}}%
\mathbf{\ }^{\alpha }h_{3}}{\mathbf{\ }^{\alpha }h_{3}}+\frac{\partial _{k}%
\mathbf{\ }^{\alpha }h_{4}}{\mathbf{\ }^{\alpha }h_{4}}\right) -\frac{_{\
_{1}x^{k}}\overset{\alpha }{\underline{\partial }}_{x^{k}}\mathbf{\ }%
^{\alpha }h_{4}^{\ast }}{2\mathbf{\ }^{\alpha }h_{4}}=0,  \notag \\
\mathbf{\ }^{\alpha }\widehat{R}_{4k} &=&\frac{\mathbf{\ }^{\alpha }h_{4}}{2%
\mathbf{\ }^{\alpha }h_{3}}\mathbf{\ }^{\alpha }n_{k}^{\ast \ast }+\left(
\frac{\mathbf{\ }^{\alpha }h_{4}}{\mathbf{\ }^{\alpha }h_{3}}\mathbf{\ }%
^{\alpha }h_{3}^{\ast }-\frac{3}{2}\mathbf{\ }^{\alpha }h_{4}^{\ast }\right)
\frac{\mathbf{\ }^{\alpha }n_{k}^{\ast }}{2\mathbf{\ }^{\alpha }h_{3}}=0,
\label{eq4}
\end{eqnarray}%
where the partial derivatives are
\begin{equation*}
\mathbf{\ }^{\alpha }a^{\bullet }=\overset{\alpha }{\underline{\partial }}%
_{1}a=_{\ _{1}x^{1}}\overset{\alpha }{\underline{\partial }}%
_{x^{1}}{}^{\alpha }a,\ \mathbf{\ }^{\alpha }a^{\prime }=\overset{\alpha }{%
\underline{\partial }}_{2}a=_{\ _{1}x^{2}}\overset{\alpha }{\underline{%
\partial }}_{x^{2}}{}^{\alpha }a,\ \ \mathbf{\ }^{\alpha }a^{\ast }=\overset{%
\alpha }{\underline{\partial }}_{v}a=_{\ _{1}v}\overset{\alpha }{\underline{%
\partial }}_{v}{}^{\alpha }a,
\end{equation*}%
 being used the left Caputo fractional derivatives (\ref{frlcb}).

Configurations with fractional Levi--Civita connection $\mathbf{\ }^{\alpha
}\nabla ,$ of type (\ref{frconstr}), can be extracted by imposing additional
constraints
\begin{eqnarray}
\mathbf{\ }^{\alpha }w_{i}^{\ast } &=&\mathbf{\ }^{\alpha }\mathbf{e}_{i}\ln
|\mathbf{\ }^{\alpha }h_{4}|,\mathbf{\ }^{\alpha }\mathbf{e}_{k}\mathbf{\ }%
^{\alpha }w_{i}=\mathbf{\ }^{\alpha }\mathbf{e}_{i}\mathbf{\ }^{\alpha
}w_{k},\ \   \notag \\
\mathbf{\ }^{\alpha }n_{i}^{\ast } &=&0,\ \overset{\alpha }{\underline{%
\partial }}_{i}\mathbf{\ }^{\alpha }n_{k}=\overset{\alpha }{\underline{%
\partial }}_{k}\mathbf{\ }^{\alpha }n_{i}.  \label{frconstr1}
\end{eqnarray}

We can construct 'non--Killing' solutions depending on all coordinates when
\begin{eqnarray}
\mathbf{\ }^{\alpha }\mathbf{g} &\mathbf{=}&\mathbf{\ }^{\alpha }g_{i}(x^{k})%
\mathbf{\ }^{\alpha }{dx^{i}\otimes \mathbf{\ }^{\alpha }dx^{i}}+\mathbf{\ }%
^{\alpha }\omega ^{2}(x^{j},v,y^{4})\mathbf{\ }^{\alpha }h_{a}(x^{k},v)%
\mathbf{\ }^{\alpha }\mathbf{e}^{a}{\otimes }\mathbf{\ }^{\alpha }\mathbf{e}%
^{a},  \notag \\
\mathbf{\ }^{\alpha }\mathbf{e}^{3} &=&\mathbf{\ }^{\alpha }dy^{3}+\mathbf{\
}^{\alpha }w_{i}(x^{k},v)\ ^{\alpha }dx^{i},\mathbf{\ }^{\alpha }\mathbf{e}%
^{4}=\mathbf{\ }^{\alpha }dy^{4}+\mathbf{\ }^{\alpha }n_{i}(x^{k},v)\mathbf{%
\ }^{\alpha }dx^{i},  \label{ansgensol}
\end{eqnarray}%
for any $\mathbf{\ }^{\alpha }\omega $ for which
\begin{equation*}
\mathbf{\ }^{\alpha }\mathbf{e}_{k}\mathbf{\ }^{\alpha }\omega =\overset{%
\alpha }{\underline{\partial }}_{k}\mathbf{\ }^{\alpha }\omega +\mathbf{\ }%
^{\alpha }w_{k}\mathbf{\ }^{\alpha }\omega ^{\ast }+\mathbf{\ }^{\alpha
}n_{k}\overset{\alpha }{\underline{\partial }}_{y^{4}}\mathbf{\ }^{\alpha
}\omega =0.
\end{equation*}

\subsubsection{Solutions with $\mathbf{\ }^{\protect\alpha }h_{3,4}^{\ast
}\neq 0$ and $\ ^{\protect\alpha }\Upsilon _{2,4}\neq 0$}

For simplicity, in this paper we show how to construct exact solution with
metrics of type (\ref{ansgensol}) when $\mathbf{\ }^{\alpha }h_{3,4}^{\ast
}\neq 0$ (in Ref. \cite{vrfrg}, there are analyzed all possibilities for
coefficients\footnote{%
by nonholonomic transforms, various classes of solutions can be transformed
from one to another}) We consider the ansatz
\begin{eqnarray}
\ \mathbf{\ }^{\alpha }\mathbf{g} &\mathbf{=}&e^{\mathbf{\ }^{\alpha }\psi
(x^{k})}\mathbf{\ }^{\alpha }{dx^{i}\otimes \mathbf{\ }^{\alpha }dx^{i}}%
+h_{3}(x^{k},v)\mathbf{\ }^{\alpha }\mathbf{e}^{3}{\otimes }\mathbf{\ }%
^{\alpha }\mathbf{e}^{3}+h_{4}(x^{k},v)\mathbf{\ }^{\alpha }\mathbf{e}^{4}{%
\otimes }\mathbf{\ }^{\alpha }\mathbf{e}^{4},  \notag \\
\mathbf{\ }^{\alpha }\mathbf{e}^{3} &=&\mathbf{\ }^{\alpha }dv+\mathbf{\ }%
^{\alpha }w_{i}(x^{k},v)\mathbf{\ }^{\alpha }dx^{i},\mathbf{\ }^{\alpha }%
\mathbf{e}^{4}=\mathbf{\ }^{\alpha }dy^{4}+\mathbf{\ }^{\alpha
}n_{i}(x^{k},v)\mathbf{\ }^{\alpha }dx^{i}  \label{genans}
\end{eqnarray}%
when
\begin{eqnarray}
\mathbf{\ }^{\alpha }\ddot{\psi}+\mathbf{\ }^{\alpha }\psi ^{\prime \prime }
&=&2\mathbf{\ }^{\alpha }\Upsilon _{4}(x^{k}),  \label{4ep1a} \\
\mathbf{\ }^{\alpha }h_{4}^{\ast } &=&2\mathbf{\ }^{\alpha }h_{3}\mathbf{\ }%
^{\alpha }h_{4}\mathbf{\ }^{\alpha }\Upsilon _{2}(x^{i},v)/\mathbf{\ }%
^{\alpha }\phi ^{\ast },  \label{4ep2a}
\end{eqnarray}%
\begin{eqnarray}
\mathbf{\ }^{\alpha }\beta \mathbf{\ }^{\alpha }w_{i}+\mathbf{\ }^{\alpha
}\alpha _{i} &=&0,  \label{4ep3a} \\
\mathbf{\ }^{\alpha }n_{i}^{\ast \ast }+\mathbf{\ }^{\alpha }\gamma \mathbf{%
\ }^{\alpha }n_{i}^{\ast } &=&0,  \label{4ep4a}
\end{eqnarray}%
\begin{eqnarray}
\mbox{where}\mathbf{\ }^{\alpha }\phi &=&\ln |\frac{\mathbf{\ }^{\alpha
}h_{4}^{\ast }}{\sqrt{|\mathbf{\ }^{\alpha }h_{3}\mathbf{\ }^{\alpha }h_{4}|}%
}|,\ \mathbf{\ }^{\alpha }\gamma =\left( \ln |\mathbf{\ }^{\alpha
}h_{4}|^{3/2}/|\mathbf{\ }^{\alpha }h_{3}|\right) ^{\ast },  \label{auxphi}
\\
\mathbf{\ }^{\alpha }\alpha _{i} &=&\mathbf{\ }^{\alpha }h_{4}^{\ast }%
\overset{\alpha }{\underline{\partial }}_{k}\ ^{\alpha }\phi ,\ \mathbf{\ }%
^{\alpha }\beta =\mathbf{\ }^{\alpha }h_{4}^{\ast }\ \mathbf{\ }^{\alpha
}\phi ^{\ast }\ .  \notag
\end{eqnarray}%
For $\mathbf{\ }^{\alpha }h_{4}^{\ast }\neq 0;\mathbf{\ }^{\alpha }\Upsilon
_{2}\neq 0,$ we have $\ ^{\alpha }\phi ^{\ast }\neq 0.$ \ The
exponent $e^{\mathbf{\ }^{\alpha }\psi (x^{k})}$ is the fractional analog of
the ''integer'' exponential functions and called the Mittag--Leffler function
$E_{\alpha }[(x-\ ^{1}x)^{\alpha }].$ For $^{\alpha }\psi (x)=E_{\alpha
}[(x-\ ^{1}x)^{\alpha }],$ we have $\overset{\alpha }{\underline{\partial }}%
_{i}E_{\alpha }=E_{\alpha },$ see (for instance) \cite{taras08}.

Choosing any nonconstant $\mathbf{\ }^{\alpha }\phi =\mathbf{\ }^{\alpha
}\phi (x^{i},v)$ as a generating function, we can construct exact solutions
of (\ref{4ep1a})--(\ref{4ep4a}). We have to solve respectively the two
dimensional fractional Laplace equation, for $\ \mathbf{\ }^{\alpha }g_{1}=\
\mathbf{\ }^{\alpha }g_{2}=e^{\ \mathbf{\ }^{\alpha }\psi (x^{k})}.$ Then we
integrate on $v,$ in order to determine $\mathbf{\ }^{\alpha }h_{3},$ $%
\mathbf{\ }^{\alpha }h_{4}$ and $\mathbf{\ }^{\alpha }n_{i},$ and solving
algebraic equations, for $\mathbf{\ }^{\alpha }w_{i}.$ We obtain (computing
consequently for a chosen $\mathbf{\ }^{\alpha }\phi (x^{k},v)$)
\begin{eqnarray}
\mathbf{\ }^{\alpha }g_{1} &=&\mathbf{\ }^{\alpha }g_{2}=e^{\mathbf{\ }%
^{\alpha }\psi (x^{k})},\mathbf{\ }^{\alpha }h_{3}=\pm \ \frac{|\mathbf{\ }%
^{\alpha }\phi ^{\ast }(x^{i},v)|}{\mathbf{\ }^{\alpha }\Upsilon _{2}},\
\label{gsol1} \\
\mathbf{\ }^{\alpha }h_{4} &=&\ \mathbf{\ }_{0}^{\alpha }h_{4}(x^{k})\pm \
2_{\ _{1}v}\overset{\alpha }{I}_{v}\frac{(\exp [2\ \mathbf{\ }^{\alpha }\phi
(x^{k},v)])^{\ast }}{\mathbf{\ }^{\alpha }\Upsilon _{2}},\   \notag \\
\mathbf{\ }^{\alpha }w_{i} &=&-\overset{\alpha }{\underline{\partial }}_{i}%
\mathbf{\ }^{\alpha }\phi /\mathbf{\ }^{\alpha }\phi ^{\ast },  \notag \\
\mathbf{\ }^{\alpha }n_{i} &=&\ _{1}^{\alpha }n_{k}\left( x^{i}\right) +\
_{2}^{\alpha }n_{k}\left( x^{i}\right) _{\ _{1}v}\overset{\alpha }{I}_{v}[%
\mathbf{\ }^{\alpha }h_{3}/(\sqrt{|\mathbf{\ }^{\alpha }h_{4}|})^{3}],
\notag
\end{eqnarray}%
where $\ \mathbf{\ }_{0}^{\alpha }h_{4}(x^{k}),\ \mathbf{\ }_{1}^{\alpha
}n_{k}\left( x^{i}\right) $ and $\ \mathbf{\ }_{2}^{\alpha }n_{k}\left(
x^{i}\right) $ are integration functions, and $_{\ _{1}v}\overset{\alpha }{I}%
_{v}$ is the fractional integral on variables $v.$

We have to constrain the coefficients (\ref{gsol1}) to satisfy the
conditions (\ref{frconstr1}) in order to construct exact solutions
for the Levi--Civita connection $\mathbf{\ }^{\alpha }\nabla .$ To
select such classes of solutions, we can fix a nonholonomic
distribution when $\ \mathbf{\ }_{2}^{\alpha }n_{k}\left(
x^{i}\right) $ $=0$ and $\ _{1}^{\alpha }n_{k}\left( x^{i}\right)
$ are any functions satisfying the conditions $\ \overset{\alpha
}{\underline{\partial }}_{i}\mathbf{\ }_{1}^{\alpha
}n_{k}\left( xj\right) =\overset{\alpha }{\underline{\partial }}_{k}\mathbf{%
\ }_{1}^{\alpha }n_{i}\left( x^{j}\right) .$ The constraints on $\mathbf{\ }%
^{\alpha }\phi (x^{k},v)$ are related to the N--connection coefficients $%
\mathbf{\ }^{\alpha }w_{i}=-\overset{\alpha }{\underline{\partial }}_{i}%
\mathbf{\ }^{\alpha }\phi /\ \mathbf{\ }^{\alpha }\phi ^{\ast }$ following
relations
\begin{eqnarray}
\left( \mathbf{\ }^{\alpha }w_{i}[\mathbf{\ }^{\alpha }\phi ]\right) ^{\ast
}+\mathbf{\ }^{\alpha }w_{i}[\mathbf{\ }^{\alpha }\phi ]\left( \mathbf{\ }%
^{\alpha }h_{4}[\mathbf{\ }^{\alpha }\phi ]\right) ^{\ast }+\overset{\alpha }%
{\underline{\partial }}_{i}\mathbf{\ }^{\alpha }h_{4}[\mathbf{\ }^{\alpha
}\phi ]=0, &&  \notag \\
\overset{\alpha }{\underline{\partial }}_{i}\mathbf{\ }^{\alpha }w_{k}[%
\mathbf{\ }^{\alpha }\phi ]=\overset{\alpha }{\underline{\partial }}_{k}\
\mathbf{\ }^{\alpha }w_{i}[\mathbf{\ }^{\alpha }\phi ], &&  \label{auxc1}
\end{eqnarray}%
where, for instance, we denoted by $\mathbf{\ }^{\alpha }h_{4}[\mathbf{\ }%
^{\alpha }\phi ]$ the functional dependence on $\mathbf{\ }^{\alpha }\phi .$
Such conditions are always satisfied for $\mathbf{\ }^{\alpha }\phi =\mathbf{%
\ }^{\alpha }\phi (v)$ or if $\mathbf{\ }^{\alpha }\phi =const$ \ when $%
\mathbf{\ }^{\alpha }w_{i}(x^{k},v)$ can be any functions as follows from (%
\ref{4ep3a}) with zero $\mathbf{\ }^{\alpha }\beta $ and $\mathbf{\ }%
^{\alpha }\alpha _{i},$ see (\ref{auxphi})).

\subsection{A geometric model for fractional Lagrange spaces}

Any solution $\mathbf{\ }^{\alpha }\mathbf{g}=\{\mathbf{\ }^{\alpha
}g_{\alpha ^{\prime }\beta ^{\prime }}(u^{\alpha ^{\prime }})\}$ of
fractional Einstein equations (\ref{fdeq}) can be parametrized in a form
derived in previous section. Using frame transforms of type $\mathbf{\ }%
^{\alpha }e_{\alpha }=e_{\ \alpha }^{\alpha ^{\prime }}\mathbf{\ }^{\alpha
}e_{\alpha ^{\prime }},$ with $\mathbf{\ }^{\alpha }\mathbf{g}_{\alpha \beta
}=e_{\ \alpha }^{\alpha ^{\prime }}e_{\ \beta }^{\beta ^{\prime }}\mathbf{\ }%
^{\alpha }g_{\alpha ^{\prime }\beta ^{\prime }},$ for any $\mathbf{\ }%
^{\alpha }\mathbf{g}_{\alpha \beta }$ (\ref{m1}), we relate the class of
such solutions, for instance, to the family of metrics of type (\ref{genans}%
). Such solutions can be related also to analogous fractional models, via
corresponding $e_{\ \alpha }^{\alpha ^{\prime }}$ to a Sasaki type metric,
see below (\ref{sasaki}).

A fractional Lagrange space is defined on a fractional tangent bundle $%
\overset{\alpha }{\underline{T}}M$ \ of \ fractional dimension $\alpha \in
(0,1)$ by a couple $\overset{\alpha }{\underline{L^{n}}}=(\overset{\alpha }{%
\underline{M}},\overset{\alpha }{L}),$ for a regular real function $\overset{%
\alpha }{L}:$ $\overset{\alpha }{\underline{T}}M\rightarrow \mathbb{R},$
when the fractional Hessian is
\begin{equation*}
\ _{L\ }\overset{\alpha }{g}_{ij}=\frac{1}{4}\left( \overset{\alpha }{%
\underline{\partial }}_{i}\overset{\alpha }{\underline{\partial }}_{j}+%
\overset{\alpha }{\underline{\partial }}_{j}\overset{\alpha }{\underline{%
\partial }}_{i}\right) \overset{\alpha }{L}\neq 0.
\end{equation*}%
Any $\overset{\alpha }{\underline{L^{n}}}$ can be associated to a prime
''integer'' Lagrange space $L^{n}.$

Let us consider values $y^{k}(\tau )=dx^{k}(\tau )/d\tau ,$ for $x(\tau )$
parametrizing smooth curves on a manifold $M$ with $\tau \in \lbrack 0,1].$
The fractional analogs of such configurations are determined by changing $\
d/d\tau $ into the fractional Caputo derivative $\ \overset{\alpha }{%
\underline{\partial }}_{\tau }=_{\ _{1}\tau }\overset{\alpha }{\underline{%
\partial }}_{\tau }$when $\ ^{\alpha }y^{k}(\tau )=\overset{\alpha }{%
\underline{\partial }}_{\tau }x^{k}(\tau ).$ Any $\overset{\alpha }{L}$
defines the fundamental geometric objects of Lagrange spaces when

\begin{enumerate}
\item the fractional Euler--Lagrange equations $\overset{\alpha }{\underline{%
\partial }}_{\tau \ }(\ _{\ _{1}y^{i}}\overset{\alpha }{\underline{\partial }%
}_{i}\overset{\alpha }{L})-_{\ _{1}x^{i}}\overset{\alpha }{\underline{%
\partial }}_{i}\overset{\alpha }{L}=0$ are equivalent to the fractional
''nonlinear geodesic'' (equivalently, semi--spray) equations $\left( \overset%
{\alpha }{\underline{\partial }}_{\tau \ }\right) ^{2}x^{k}+2\overset{\alpha
}{G^{k}}(x,\ ^{\alpha }y)=0,$ where
\begin{equation*}
\overset{\alpha }{G^{k}}=\frac{1}{4}\ \ _{L\ }\overset{\alpha }{g^{kj}}\left[
y^{j}\ _{\ _{1}y^{j}}\overset{\alpha }{\underline{\partial }}_{j}\ \left(
_{\ _{1}x^{i}}\overset{\alpha }{\underline{\partial }}_{i}\overset{\alpha }{L%
}\right) -\ _{\ _{1}x^{i}}\overset{\alpha }{\underline{\partial }}_{i}%
\overset{\alpha }{L}\right]
\end{equation*}
defines the canonical N--connection $\ _{L}^{\alpha }N_{j}^{a}=\ _{\
_{1}y^{j}}\overset{\alpha }{\underline{\partial }}_{j}\overset{\alpha }{G^{k}%
}(x,\ ^{\alpha }y);$

\item the canonical (Sasaki type) metric structure is
\begin{equation}
\ \ _{L}\overset{\alpha }{\mathbf{g}}=\ _{L}^{\alpha }g_{kj}(x,y)\ ^{\alpha
}e^{k}\otimes \ ^{\alpha }e^{j}+\ _{L}^{\alpha }g_{cb}(x,y)\ _{L}^{\alpha }%
\mathbf{e}^{c}\otimes \ _{L}^{\alpha }\mathbf{e}^{b},  \label{sasaki}
\end{equation}%
where the frame structure $\ _{L}^{\alpha }\mathbf{e}_{\nu }=(\ _{L}^{\alpha
}\mathbf{e}_{i},e_{a})$ is linear on $\ _{L}^{\alpha }N_{j}^{a};$

\item the canonical metrical d--connection
\begin{equation*}
\ _{c}^{\alpha }\mathbf{D}=(h\ _{c}^{\alpha }D,v\ _{c}^{\alpha }D)=\{\
_{c}^{\alpha }\mathbf{\Gamma }_{\ \alpha \beta }^{\gamma }=(\ ^{\alpha }%
\widehat{L}_{\ jk}^{i},\ ^{\alpha }\widehat{C}_{jc}^{i})\}\
\end{equation*}%
is a metric compatible, $\ _{c}^{\alpha }\mathbf{D}\ \left( \ \ _{L}\overset{%
\alpha }{\mathbf{g}}\right) =0,$ for
\begin{equation*}
\ _{c}^{\alpha }\mathbf{\Gamma }_{\ j}^{i}=\ _{c}^{\alpha }\mathbf{\Gamma }%
_{\ j\gamma }^{i}\ _{L}^{\alpha }\mathbf{e}^{\gamma }=\widehat{L}_{\
jk}^{i}e^{k}+\widehat{C}_{jc}^{i}\ _{L}^{\alpha }\mathbf{e}^{c},
\end{equation*}%
with $\widehat{L}_{\ jk}^{i}=\widehat{L}_{\ bk}^{a},\widehat{C}_{jc}^{i}=%
\widehat{C}_{bc}^{a}$ in $\ \ _{c}^{\alpha }\mathbf{\Gamma }_{\ b}^{a}=\
_{c}^{\alpha }\mathbf{\Gamma }_{\ b\gamma }^{a}\ _{L}^{\alpha }\mathbf{e}%
^{\gamma }=\widehat{L}_{\ bk}^{a}e^{k}+\widehat{C}_{bc}^{a}\ _{L}^{\alpha }%
\mathbf{e}^{c},$\footnote{%
for integer dimensions, we contract ''horizontal'' and ''vertical'' indices
following the rule: $i=1$ is $a=n+1;$ $i=2$ is $a=n+2;$ ... $i=n$ is $a=n+n"$%
} and generalized Christoffel indices
\begin{eqnarray*}
\ ^{\alpha }\widehat{L}_{jk}^{i} &=&\frac{1}{2}\ _{L}^{\alpha }g^{ir}\left(
\ _{L}^{\alpha }\mathbf{e}_{k}\ _{L}^{\alpha }g_{jr}+\ _{L}^{\alpha }\mathbf{%
e}_{j}\ _{L}^{\alpha }g_{kr}-\ _{L}^{\alpha }\mathbf{e}_{r}\ _{L}^{\alpha
}g_{jk}\right) , \\
\ \ ^{\alpha }\widehat{C}_{bc}^{a} &=&\frac{1}{2}\ _{L}^{\alpha
}g^{ad}\left( \ ^{\alpha }e_{c}\ _{L}^{\alpha }g_{bd}+\ ^{\alpha }e_{c}\
_{L}^{\alpha }g_{cd}-\ ^{\alpha }e_{d}\ _{L}^{\alpha }g_{bc}\right) .
\end{eqnarray*}
\end{enumerate}

\section{Nonholonomic Fractional Curve Flows}

We formulate a model of geometry of curve flows adapted to a N--connection
structure on a fractional manifold $\overset{\alpha }{\mathbf{V}}\mathbf{.}$

\subsection{Non--stretching and N--adapted fractional curve flows}

A canonical d--connection operator $\ ^{\alpha }\widehat{\mathbf{D}}$ (\ref%
{candcon}) acts in the form%
\begin{equation}
\ ^{\alpha }\widehat{\mathbf{D}}_{\mathbf{X}}\ ^{\alpha }\mathbf{e}_{\alpha
}=(\mathbf{X\rfloor }\ ^{\alpha }\mathbf{\Gamma }_{\alpha \ }^{\ \gamma })\
^{\alpha }\mathbf{e}_{\gamma }\mbox{ and }\ ^{\alpha }\widehat{\mathbf{D}}_{%
\mathbf{Y}}\ ^{\alpha }\mathbf{e}_{\alpha }=(\mathbf{Y\rfloor \ }^{\alpha }%
\mathbf{\Gamma }_{\alpha \ }^{\ \gamma })\ ^{\alpha }\mathbf{e}_{\gamma },
\label{part01}
\end{equation}%
where ''$\mathbf{\rfloor "}$ denotes the interior product. The value $\
^{\alpha }\widehat{\mathbf{D}}_{\mathbf{X}}=\mathbf{X}^{\beta }\ ^{\alpha }%
\widehat{\mathbf{D}}_{\beta }$ is a covariant derivation operator along
curve $\gamma (\tau ,\mathbf{l}).$ It is convenient to fix the N--adapted
frame to be parallel to curve $\gamma (\mathbf{l})$ adapted in the form
\begin{eqnarray}
\ ^{\alpha }e^{1} &\doteqdot &h\mathbf{X,}\mbox{ for }i=1,\mbox{ and }\
^{\alpha }e^{\widehat{i}},\mbox{ where }h\ ^{\alpha }\mathbf{g(}h\mathbf{X,}%
\ ^{\alpha }e^{\widehat{i}}\mathbf{)=}0,  \label{curvframe} \\
\ ^{\alpha }\mathbf{e}^{n+1} &\doteqdot &v\mathbf{X,}\mbox{ for }a=n+1,%
\mbox{ and }\ ^{\alpha }\mathbf{e}^{\widehat{a}},\mbox{ where }v\ ^{\alpha }%
\mathbf{g(}v\mathbf{X,}\ ^{\alpha }\mathbf{\mathbf{e}}^{\widehat{a}}\mathbf{%
)=}0,  \notag
\end{eqnarray}%
for $\widehat{i}=2,3,...n$ and $\widehat{a}=n+2,n+3,...,n+m.$ The covariant
derivative of each ''normal'' d--vectors $\ ^{\alpha }\mathbf{e}^{\widehat{%
\alpha }}$ results into d--vectors adapted to $\gamma (\tau ,\mathbf{l}),$
\begin{eqnarray}
\ ^{\alpha }\widehat{\mathbf{D}}_{\mathbf{X}}\ ^{\alpha }e^{\widehat{i}} &%
\mathbf{=}&\mathbf{-}\rho ^{\widehat{i}}\mathbf{(}u\mathbf{)\ X}\mbox{ and }%
\ ^{\alpha }\widehat{\mathbf{D}}_{h\mathbf{X}}h\mathbf{X}=\rho ^{\widehat{i}}%
\mathbf{(}u\mathbf{)\ }\ ^{\alpha }\mathbf{\mathbf{e}}_{\widehat{i}},
\label{part02} \\
\ ^{\alpha }\widehat{\mathbf{D}}_{\mathbf{X}}\ ^{\alpha }\mathbf{\mathbf{e}}%
^{\widehat{a}} &\mathbf{=}&\mathbf{-}\rho ^{\widehat{a}}\mathbf{(}u\mathbf{%
)\ X}\mbox{ and }\ ^{\alpha }\widehat{\mathbf{D}}_{v\mathbf{X}}v\mathbf{X}%
=\rho ^{\widehat{a}}\mathbf{(}u\mathbf{)\ }\ ^{\alpha }e_{\widehat{a}},
\notag
\end{eqnarray}%
which holds for certain classes of functions $\rho ^{\widehat{i}}\mathbf{(}u%
\mathbf{)}$ and $\rho ^{\widehat{a}}\mathbf{(}u\mathbf{).}$ The formulas (%
\ref{part01}) and (\ref{part02}) are distinguished into h-- and
v--components for $\mathbf{X=}h\mathbf{X}+v\mathbf{X}$ and $\ ^{\alpha }%
\widehat{\mathbf{D}}\mathbf{=(}h\mathbf{D},v\mathbf{D)}$ for $\ ^{\alpha }%
\widehat{\mathbf{D}}=\{\ ^{\alpha }\widehat{\mathbf{\Gamma }}_{\
\alpha \beta }^{\gamma }\},h\mathbf{D}=\{\ ^{\alpha }\widehat{L}_{jk}^{i},\
^{\alpha }\widehat{L}_{bk}^{a}\}$ and $v\mathbf{D=\{}\ ^{\alpha }\widehat{C}%
_{jc}^{i},\ ^{\alpha }\widehat{C}_{bc}^{a}\}.$

A non--stretching curve $\gamma (\tau ,\mathbf{l})$ on $\overset{\alpha }{%
\mathbf{V}}\mathbf{,}$ where $\tau $ is a real parameter and $\mathbf{l}$ is
the arclength of the curve on $\overset{\alpha }{\mathbf{V}}\mathbf{,}$ is
defined with such evolution d--vector $\mathbf{Y}=\gamma _{\tau }$ and
tangent d--vector $\mathbf{X}=\gamma _{\mathbf{l}}$ that $\ ^{\alpha }%
\mathbf{g(X,X)=}1\mathbf{.}$ Such a curve $\gamma (\tau ,\mathbf{l})$ swept
out a two--dimensional surface in $\underline{T}_{\gamma (\tau ,\mathbf{l})}%
\overset{\alpha }{\mathbf{V}}\subset \underline{T}\overset{\alpha }{\mathbf{V%
}}\mathbf{.}$ Along $\gamma (\mathbf{l}),$ we can move differential forms in
a parallel N--adapted form. For instance, $\ \ ^{\alpha }\widehat{\mathbf{%
\Gamma }}_{\ \mathbf{X}}^{\alpha \beta }\doteqdot \mathbf{X\rfloor }\
^{\alpha }\mathbf{\widehat{\mathbf{\Gamma }}}_{\ }^{\alpha \beta }.$ Such
fractional spaces can be characterized algebraically if we perform a frame
transform preserving the decomposition (\ref{whit}) to an orthonormalized
basis $\ ^{\alpha }\mathbf{e}_{\alpha ^{\prime }},$ when
\begin{equation}
\ ^{\alpha }\mathbf{e}_{\alpha }\rightarrow e_{\alpha }^{\ \alpha ^{\prime
}}(u)\ \ ^{\alpha }\mathbf{e}_{\alpha ^{\prime }},  \label{orthbas}
\end{equation}%
is an orthonormal d--basis. In this case, the coefficients of the d--metric (%
\ref{m1}) transform into the (pseudo) Euclidean one\textbf{\ }$\ ^{\alpha }%
\mathbf{g}_{\alpha ^{\prime }\beta ^{\prime }}=\eta _{\alpha ^{\prime }\beta
^{\prime }}$ by encoding the fractional configuration into the structure of
local d--bases. We obtain two skew matrices%
\begin{equation*}
\ ^{\alpha }\widehat{\mathbf{\Gamma }}_{h\mathbf{X}}^{i^{\prime }j^{\prime
}}\doteqdot h\mathbf{X\rfloor }\ ^{\alpha }\widehat{\mathbf{\Gamma }}_{\
}^{i^{\prime }j^{\prime }}=2\ \ ^{\alpha }e_{h\mathbf{X}}^{[i^{\prime }}\
\rho ^{j^{\prime }]}\mbox{ and }\ \ ^{\alpha }\widehat{\mathbf{\Gamma }}_{v%
\mathbf{X}}^{a^{\prime }b^{\prime }}\doteqdot v\mathbf{X\rfloor }\ ^{\alpha }%
\widehat{\mathbf{\Gamma }}_{\ }^{a^{\prime }b^{\prime }}=2\mathbf{\ e}_{v%
\mathbf{X}}^{[a^{\prime }}\ \rho ^{b^{\prime }]},
\end{equation*}%
\begin{eqnarray*}
\ e_{h\mathbf{X}}^{i^{\prime }}&\doteqdot& g(h\mathbf{X,}e^{i^{\prime }})=[1,%
\underbrace{0,\ldots ,0}_{n-1}]\mbox{ and }\ e_{v\mathbf{X}}^{a^{\prime
}}\doteqdot h(v\mathbf{X,}e^{a^{\prime }})=[1,\underbrace{0,\ldots ,0}%
_{m-1}], \\
\ ^{\alpha }\widehat{\mathbf{\Gamma }}_{h\mathbf{X\,}i^{\prime }}^{\qquad
j^{\prime }}&=&\left[
\begin{array}{cc}
0 & \rho ^{j^{\prime }} \\
-\rho _{i^{\prime }} & \mathbf{0}_{[h]}%
\end{array}%
\right] \mbox{ and }\ ^{\alpha }\widehat{\mathbf{\Gamma }}_{v\mathbf{X\,}%
a^{\prime }}^{\qquad b^{\prime }}=\left[
\begin{array}{cc}
0 & \rho ^{b^{\prime }} \\
-\rho _{a^{\prime }} & \mathbf{0}_{[v]}%
\end{array}%
\right]
\end{eqnarray*}%
with $\mathbf{0}_{[h]}$ and $\mathbf{0}_{[v]}$ being respectively $%
(n-1)\times (n-1)$ and $(m-1)\times (m-1)$ matrices.\footnote{%
The above presented row--matrices and skew--matrices show that locally an
N--anholonomic fractiona manifold $\overset{\alpha }{\mathbf{V}}$ are
related to prime spaces of integer--dimension $n+m.$ With respect to
distinguished orthonormalized frames the constrained fractional dynamics is
characterized algebraically by couples of unit vectors in $\mathbb{R}^{n}$
and $\mathbb{R}^{m}$ preserved respectively by the $SO(n-1)$ and $SO(m-1)$
rotation subgroups of the local N--adapted frame structure group $%
SO(n)\oplus SO(m).$ The connection matrices $\ ^{\alpha }\widehat{\mathbf{%
\Gamma }}_{h\mathbf{X\,}i^{\prime }}^{\qquad j^{\prime }}$ and $\ ^{\alpha }%
\widehat{\mathbf{\Gamma }}_{v\mathbf{X\,}a^{\prime }}^{\qquad b^{\prime }}$
belong to the orthogonal complements of the corresponding Lie subalgebras
and algebras, $\mathfrak{so}(n-1)\subset \mathfrak{so}(n)$ and $\mathfrak{so}%
(m-1)\subset \mathfrak{so}(m).$}

The torsion and curvature tensors (\ref{tors}) can be written in
orthonormalized component form with respect to (\ref{curvframe}) mapped into
a distinguished orthonormalized dual frame (\ref{orthbas}),%
\begin{eqnarray}
\ ^{\alpha }\widehat{\mathcal{T}}^{\alpha ^{\prime }}&\doteqdot& \ ^{\alpha }%
\widehat{\mathbf{D}}_{\mathbf{X}}\ ^{\alpha }\mathbf{e}_{\mathbf{Y}}^{\alpha
^{\prime }}-\ ^{\alpha }\widehat{\mathbf{D}}_{\mathbf{Y}}\ ^{\alpha }\mathbf{%
e}_{\mathbf{X}}^{\alpha ^{\prime }}+\ ^{\alpha }\mathbf{e}_{\mathbf{Y}%
}^{\beta ^{\prime }}\ ^{\alpha }\widehat{\mathbf{\Gamma }}_{\mathbf{X}\beta
^{\prime }}^{\quad \alpha ^{\prime }}-\ ^{\alpha }\mathbf{e}_{\mathbf{X}%
}^{\beta ^{\prime }}\ ^{\alpha }\widehat{\mathbf{\Gamma }}_{\mathbf{Y}\beta
^{\prime }}^{\quad \alpha ^{\prime }}  \label{mtors} \\
&&\mbox{ and} \ ^{\alpha }\widehat{\mathcal{R}}_{\beta ^{\prime }}^{\;\alpha
^{\prime }}(\mathbf{X,Y})=\ ^{\alpha }\widehat{\mathbf{D}}_{\mathbf{Y}}\
^{\alpha }\widehat{\mathbf{\Gamma }}_{\mathbf{X}\beta ^{\prime }}^{\quad
\alpha ^{\prime }}-\ ^{\alpha }\widehat{\mathbf{D}}_{\mathbf{X}}\ ^{\alpha }%
\widehat{\mathbf{\Gamma }}_{\mathbf{Y}\beta ^{\prime }}^{\quad \alpha
^{\prime }}  \notag \\
&&+\ ^{\alpha }\widehat{\mathbf{\Gamma }}_{\mathbf{Y}\beta ^{\prime
}}^{\quad \gamma ^{\prime }}\ ^{\alpha }\widehat{\mathbf{\Gamma }}_{\mathbf{X%
}\gamma ^{\prime }}^{\quad \alpha ^{\prime }}-\ ^{\alpha }\widehat{\mathbf{%
\Gamma }}_{\mathbf{X}\beta ^{\prime }}^{\quad \gamma ^{\prime }}\ ^{\alpha }%
\widehat{\mathbf{\Gamma }}_{\mathbf{Y}\gamma ^{\prime }}^{\quad \alpha
^{\prime }}.  \label{mcurv}
\end{eqnarray}%
The values $\ ^{\alpha }\mathbf{e}_{\mathbf{Y}}^{\alpha ^{\prime }}\doteqdot
\ ^{\alpha }\mathbf{g}(\mathbf{Y},\ ^{\alpha }\mathbf{e}^{\alpha ^{\prime
}}), \ ^{\alpha }\widehat{\mathbf{\Gamma }}_{\mathbf{Y}\beta ^{\prime
}}^{\quad \alpha ^{\prime }}\doteqdot \mathbf{Y\rfloor }\ ^{\alpha }\widehat{%
\mathbf{\Gamma }}_{\beta ^{\prime }}^{\;\alpha ^{\prime }}=\ ^{\alpha }%
\mathbf{g}(\ ^{\alpha }\mathbf{e}^{\alpha ^{\prime }},\ ^{\alpha }\widehat{%
\mathbf{D}}_{\mathbf{Y}}\ ^{\alpha }\mathbf{e}_{\beta ^{\prime }})$ define
respectively the N--adapted orthonormalized frame row--matrix and the
canonical d--connection skew--matrix in the flow directs, and \newline
$\ ^{\alpha }\widehat{\mathcal{R}}_{\beta ^{\prime }}^{\;\alpha ^{\prime }}(%
\mathbf{X,Y})\doteqdot \ ^{\alpha }\mathbf{g}(\ ^{\alpha }\mathbf{e}^{\alpha
^{\prime }},[\ ^{\alpha }\widehat{\mathbf{D}}_{\mathbf{X}},$ $\ ^{\alpha }%
\widehat{\mathbf{D}}_{\mathbf{Y}}]\ ^{\alpha }\mathbf{e}_{\beta ^{\prime }})$
is the curvature matrix. 

\subsection{N--anholonomic fractional manifolds with constant matrix curvature}

The geometry of integer dimensional Einstein and Lagrange--Finsler spa\-ces
can be encoded into bi--Hamilton structures and associated solitonic
hi\-erarichies \cite{vacap,vanco}. \ The goal of this section is to show that
there is a geometric background for extending the constructions to the case
of fractional spaces. We shall elaborate the concept of fractional
N--anholonomic Klein space which in \cite{bv4} will be applied for
constructing fractional solitonic hierarchies.

\subsubsection{Symmetric fractional nonholonomic manifolds}

For trivial N--connection curvature and torsion but constant matrix
curvature on spaces of integer dimension, we get a holonomic Riemannian
manifold and the equations (\ref{mtors}) and (\ref{mcurv}) directly encode a
bi--Hamiltonian structure \cite{saw,anc2}. A well known class of Riemannian
manifolds for which the frame curvature matrix constant consists of the
symmetric spaces $M=G/H$ for compact semisimple Lie groups $G\supset H.$ A
complete classification and summary of main results on such integer
dimension spaces are given in Ref. \cite{helgason}. Using the Caputo
partial derivative, such a classification can be provided for fractional
spaces of constant matrix curvature. The derived algebraic classification is
that for a used ''prime'' integer dimension space but the differential and
integral calculus are those for fractional nonholonomic distributions.

Algebraically, our aim is to solder in a canonic way the horizontal and
vertical symmetric Riemannian spaces of dimension $n$ and $m$ with a (total)
symmetric Riemannian space $V$ of dimension $n+m,$ when $V=G/SO(n+m)$ with
the isotropy group $H=SO(n+m)\supset O(n+m)$ and $G=SO(n+m+1).$ The Caputo
fractional derivative is encoded for constructing tangent spaces and related
geometric objects. For the just mentioned horizontal, vertical and total
symmetric Riemannian spaces one exists natural settings to Klein geometry. A
prime fractional metric tensor $h\ ^{\alpha }g=\{\ ^{\alpha }\mathring{g}%
_{ij}\}$ on $h\ ^{\alpha }\mathbf{V}$ is defined by the Cartan--Killing
inner product $<\cdot ,\cdot >_{h}$ on $\underline{T}_{x}hG\simeq h\mathfrak{%
g}$ restricted to the Lie algebra quotient spaces $h\mathfrak{p=}h\mathfrak{%
g/}h\mathfrak{h,}$ with $\underline{T}_{x}hH\simeq h\mathfrak{h,}$ where $h%
\mathfrak{g=}h\mathfrak{h}\oplus h\mathfrak{p}$ is stated such that there is
an involutive automorphism of $hG$ under $hH$ is fixed, i.e. $[h\mathfrak{h,}%
h\mathfrak{p]}\subseteq $ $h\mathfrak{p}$ and $[h\mathfrak{p,}h\mathfrak{p]}%
\subseteq h\mathfrak{h.}$ In a similar form, we can define the group spaces
and related inner products and\ Lie algebras,%
\begin{eqnarray*}
\mbox{for\ }vg &=&\{\mathring{h}_{ab}\},\;<\cdot ,\cdot >_{v},\;\underline{T}%
_{y}vG\simeq v\mathfrak{g,\;}v\mathfrak{p=}v\mathfrak{g/}v\mathfrak{h,}%
\mbox{ with } \\
\underline{T}_{y}vH &\simeq &v\mathfrak{h,}v\mathfrak{g=}v\mathfrak{h}\oplus
v\mathfrak{p,}\mbox{where }\mathfrak{\;}[v\mathfrak{h,}v\mathfrak{p]}%
\subseteq v\mathfrak{p,\;}[v\mathfrak{p,}v\mathfrak{p]}\subseteq v\mathfrak{%
h;} \\
&& \\
\mbox{for\ }\mathbf{g} &=&\{\mathring{g}_{\alpha \beta }\},\;<\cdot ,\cdot
>_{\mathbf{g}},\;\underline{T}_{(x,y)}G\simeq \mathfrak{g,\;p=g/h,}%
\mbox{
with } \\
\underline{T}_{(x,y)}H &\simeq &\mathfrak{h,g=h}\oplus \mathfrak{p,}%
\mbox{where }\mathfrak{\;}[\mathfrak{h,p]}\subseteq \mathfrak{p,\;}[%
\mathfrak{p,p]}\subseteq \mathfrak{h.}
\end{eqnarray*}%
Similar formulas in \cite{vacap,vanco} are for usual partial derivatives
with not underlined symbols $T.$ We parametrize the metric structure with
constant coefficients on $V=G/SO(n+m)$ and fractional differentials in the
form%
\begin{equation*}
\ ^{\alpha }\mathring{g}=\ ^{\alpha }\mathring{g}_{\underline{\gamma }%
\underline{\beta }}(du^{\underline{\gamma }})^{\alpha }\otimes (du^{%
\underline{\beta }})^{\alpha },
\end{equation*}%
where the coefficients of a fractional metric of type (\ref{fmcf}) are
parametrized in the form
\begin{equation*}
\ ^{\alpha }\mathring{g}_{\alpha \beta }=\left[
\begin{array}{cc}
\ ^{\alpha }\mathring{g}_{ij}+\ ^{\alpha }\mathring{N}_{i}^{a}\ ^{\alpha
}N_{j}^{b}\ ^{\alpha }\mathring{h}_{ab} & \ ^{\alpha }\mathring{N}_{j}^{e}\
^{\alpha }\mathring{h}_{ae} \\
\ ^{\alpha }\mathring{N}_{i}^{e}\ ^{\alpha }\mathring{h}_{be} & \ ^{\alpha }%
\mathring{h}_{ab}%
\end{array}%
\right]
\end{equation*}%
with trivial, constant, N--connection coefficients computed $\ ^{\alpha }%
\mathring{N}_{j}^{e}=\ ^{\alpha }\mathring{h}^{eb}\ ^{\alpha }\mathring{g}%
_{jb}$ for any given sets $\ ^{\alpha }\mathring{h}^{eb}$ and $\ ^{\alpha }%
\mathring{g}_{jb},$ i.e. from the inverse metrics coefficients defined
respectively on $hG=SO(n+1).$ We can also define an equivalent d--metric
structure of type (\ref{m1})
\begin{eqnarray*}
\ ^{\alpha }\mathbf{\mathring{g}} &=&\ ^{\alpha }\ \mathring{g}_{ij}\
^{\alpha }\ e^{i}\otimes \ ^{\alpha }e^{j}+\ \ ^{\alpha }\mathring{h}_{ab}\
\ ^{\alpha }\mathbf{\mathring{e}}^{a}\otimes \ ^{\alpha }\mathbf{\mathring{e}%
}^{b}, \\
\ ^{\alpha }e^{i} &=&(dx^{i})^{\alpha },\ \;\ ^{\alpha }\mathbf{\mathring{e}}%
^{a}=(dy^{a})^{\alpha }+\ ^{\alpha }\mathring{N}_{i}^{a}(dx^{i})^{\alpha }.
\end{eqnarray*}

We note that for integer dimensions such trivial parametrizations define
algebraic classifications of \ symmetric Riemannian spaces of dimension $n+m$
with constant matrix curvature admitting splitting (by certain algebraic
constraints) into symmetric Riemannian subspaces of dimension $n$ and $m,$
also both with constant matrix curvature and introducing the concept of
N--anholonomic Riemannian space of type $\ ^{\alpha }\mathbf{\mathring{V}}%
=[hG=SO(n+1),$ $vG=SO(m+1),\;\ ^{\alpha }\mathring{N}_{i}^{e}].$ Such spaces
of constant distinguished curvature are constructed as trivially
N--anholonomic group spaces which possess a Lie d--algebra symmetry $%
\mathfrak{so}_{\mathring{N}}(n+m)\doteqdot \mathfrak{so}(n)\oplus \mathfrak{%
so}(m).$

A fractional generalization of constructions is to consider nonholonomic
distributions on $V=G/SO(n+m)$ defined locally by arbitrary N--connection
coefficients $\ ^{\alpha }N_{i}^{a}(x,y)$ with nonvanishing $\ ^{\alpha
}W_{\alpha \beta }^{\gamma }$ and $\ ^{\alpha }\Omega _{ij}^{a}$ but with
constant d--metric coefficients when the fractional metric $\ ^{\alpha }%
\mathbf{g}$ is of type $\ ^{\alpha }\mathbf{g}_{\alpha ^{\prime }\beta
^{\prime }}=[\ ^{\alpha }g_{i^{\prime }j^{\prime }},\ ^{\alpha }h_{a^{\prime
}b^{\prime }}]$ (\ref{m1}) with constant coefficients $\ ^{\alpha
}g_{i^{\prime }j^{\prime }}=\ _{0}^{\alpha }g_{i^{\prime }j^{\prime }}=\
^{\alpha }\ \mathring{g}_{ij}$ and $\ ^{\alpha }h_{a^{\prime }b^{\prime }}=\
_{0}^{\alpha }h_{a^{\prime }b^{\prime }}=\ ^{\alpha }\mathring{h}_{ab}$ (in
this section \ induced by the corresponding Lie d--algebra structure $%
\mathfrak{so}_{\mathring{N}}(n+m)).$ Such spaces transform into
N--anholonomic fractional manifolds $\ ^{\alpha }\mathbf{\mathring{V}}_{%
\mathbf{N}}=[hG=SO(n+1),$ $vG=SO(m+1),\;\ ^{\alpha }N_{i}^{e}]$ \ with
nontrivial N--connection curvature and induced d--torsion coefficients of
the canonical d--connection\footnote{%
see formulas (\ref{tors}) computed for constant d--metric coefficients and
the canonical d--connection coefficients in (\ref{candcon})}.

\subsubsection{Fractional N--anholonomic Klein spaces}

We can characterize curve flows (both in integer and fractional dimensions)
\ by two Hamiltonian variables given by the principal normals $\;^{h}\nu $
and $\;^{v}\nu ,$ respectively, in the horizontal and vertical subspaces,
defined by the canonical d--connection $\ ^{\alpha }\widehat{\mathbf{D}}=(h%
\mathbf{D},v\mathbf{D}),$ $\;^{h}\nu \doteqdot \ ^{\alpha }\widehat{\mathbf{D%
}}_{h\mathbf{X}}h\mathbf{X}=\nu ^{\widehat{i}}\ ^{\alpha }\mathbf{\mathbf{e}}%
_{\widehat{i}}$ and $\ ^{v}\nu \doteqdot \ ^{\alpha }\widehat{\mathbf{D}}_{v%
\mathbf{X}}v\mathbf{X}=\nu ^{\widehat{a}}\ ^{\alpha }e_{\widehat{a}},$ see
formulas (\ref{curvframe}) and (\ref{part02}). This normal fractional
d--vector $\ ^{\alpha }\mathbf{v}=(\;^{h}\nu ,$ $\;^{v}\nu ),$ with
components of type $\ ^{\alpha }\mathbf{\nu }^{\alpha }=(\nu ^{i},$ $\;\nu
^{a})=(\nu ^{1},$ $\nu ^{\widehat{i}},\nu ^{n+1},\nu ^{\widehat{a}}),$
encoding the Caputo fractional derivative is in the tangent direction of
curve $\gamma .$ It can be also considered the principal normal d--vector $\
^{\alpha }\mathbf{\varpi }=(\;^{h}\varpi ,\;^{v}\varpi )$ with components of
type $\ ^{\alpha }\mathbf{\varpi }^{\alpha }=(\varpi ^{i},$ $\;\varpi
^{a})=(\varpi ^{1},\varpi ^{\widehat{i}},\varpi ^{n+1},\varpi ^{\widehat{a}%
}) $ in the flow direction, with
\begin{equation*}
\;^{h}\varpi \doteqdot \ ^{\alpha }\widehat{\mathbf{D}}_{h\mathbf{Y}}h%
\mathbf{X=}\varpi ^{\widehat{i}}\ ^{\alpha }\mathbf{\mathbf{e}}_{\widehat{i}%
},\;^{v}\varpi \doteqdot \ ^{\alpha }\widehat{\mathbf{D}}_{v\mathbf{Y}}v%
\mathbf{X}=\varpi ^{\widehat{a}}\ ^{\alpha }e_{\widehat{a}},
\end{equation*}%
representing a fractional Hamiltonian d--covector field. We argue that the
normal part of the flow d--vector $\ \mathbf{h}_{\perp }\doteqdot \mathbf{Y}%
_{\perp }=h^{\widehat{i}}\ ^{\alpha }\mathbf{\mathbf{e}}_{\widehat{i}}+h^{%
\widehat{a}}\ ^{\alpha }e_{\widehat{a}}$ represents a fractional Hamiltonian
d--vector field and use parallel N--adapted frames $\ ^{\alpha }\mathbf{e}%
_{\alpha ^{\prime }}=(\ ^{\alpha }\mathbf{e}_{i^{\prime }},\ ^{\alpha
}e_{a^{\prime }})$ when the h--variables $\nu ^{\widehat{i^{\prime }}},$ $%
\varpi ^{\widehat{i^{\prime }}},h^{\widehat{i^{\prime }}}$ are respectively
encoded in the top row of the horizontal canonical d--connection matrices $\
^{\alpha }\widehat{\mathbf{\Gamma }}_{h\mathbf{X\,}i^{\prime }}^{\qquad
j^{\prime }}$ and $\ ^{\alpha }\widehat{\mathbf{\Gamma }}_{h\mathbf{Y\,}%
i^{\prime }}^{\qquad j^{\prime }}$ and in the row matrix $\left( \mathbf{e}_{%
\mathbf{Y}}^{i^{\prime }}\right) _{\perp }\doteqdot \mathbf{e}_{\mathbf{Y}%
}^{i^{\prime }}-g_{\parallel }\;\mathbf{e}_{\mathbf{X}}^{i^{\prime }},$
where $g_{\parallel }\doteqdot \ ^{\alpha }g(h\mathbf{Y,}h\mathbf{X})$ is
the tangential h--part of the fractional flow d--vector.\footnote{%
It is possible to encode v--variables $\nu ^{\widehat{a^{\prime }}},\varpi ^{%
\widehat{a^{\prime }}},h^{\widehat{a^{\prime }}}$ in the top row of the
vertical canonical d--connection matrices \ $\ ^{\alpha }\widehat{\mathbf{%
\Gamma }}_{v\mathbf{X\,}a^{\prime }}^{\qquad b^{\prime }}$ and $\ ^{\alpha }%
\widehat{\mathbf{\Gamma }}_{v\mathbf{Y\,}a^{\prime }}^{\qquad b^{\prime }}$
and in the row matrix $\left( \mathbf{e}_{\mathbf{Y}}^{a^{\prime }}\right)
_{\perp }\doteqdot \mathbf{e}_{\mathbf{Y}}^{a^{\prime }}-h_{\parallel }\;%
\mathbf{e}_{\mathbf{X}}^{a^{\prime }}$ where $h_{\parallel }\doteqdot \
^{\alpha }h(v\mathbf{Y,}v\mathbf{X})$ is the tangential v--part of the flow
d--vector. In a compact form of notations, we shall write $\mathbf{v}%
^{\alpha ^{\prime }}$ and $\mathbf{\varpi }^{\alpha ^{\prime }}$ where the
primed small Greek indices $\alpha ^{\prime },\beta ^{\prime },...$ will
denote both N--adapted and then orthonormalized components of geometric
objects (d--vectors, d--covectors, d--tensors, d--groups, d--algebras,
d--matrices) admitting further decompositions into h-- and v--components.}

A N--connection structure (in particular, a fractional Lagrangian) induces a
N--anholonomic Klein space stated by two left--invariant $h\mathfrak{g}$--
and $v\mathfrak{g}$--valued Maurer--Cartan form on the Lie d--group $\mathbf{%
G}=(h\mathbf{G},v\mathbf{G})$ is identified with the zero--curvature
canonical d--connection 1--form $\ _{\mathbf{G}}^{\alpha }\widehat{\mathbf{%
\Gamma }}=\{\;\ _{\mathbf{G}}^{\alpha }\widehat{\mathbf{\Gamma }}_{\ \beta
^{\prime }}^{\alpha ^{\prime }}\},$
\begin{equation*}
\;\;\ _{\mathbf{G}}^{\alpha }\widehat{\mathbf{\Gamma }}_{\ \beta ^{\prime
}}^{\alpha ^{\prime }}=\;\;\ _{\mathbf{G}}^{\alpha }\widehat{\mathbf{\Gamma }%
}_{\ \beta ^{\prime }\gamma ^{\prime }}^{\alpha ^{\prime }}\mathbf{e}%
^{\gamma ^{\prime }}=\ _{h\mathbf{G}}^{\alpha }\widehat{L}_{\;j^{\prime
}k^{\prime }}^{i^{\prime }}\ ^{\alpha }\mathbf{e}^{k^{\prime }}+\;\ _{v%
\mathbf{G}}^{\alpha }\widehat{C}_{\;j^{\prime }k^{\prime }}^{i^{\prime }}\
^{\alpha }e^{k^{\prime }}.
\end{equation*}%
For $n=m,$ and canonical d--objects (N--connection, d--metric,
d--connection, ...) derived from (\ref{m1}), and any N--anholonomic space
with constant d--curvatures, the Cartan d--connection transform just in the
canonical d--connection (\ref{candcon}). Using the Lie d--algebra
decompositions $\mathfrak{g}=h\mathfrak{g}\oplus v\mathfrak{g,}$ for the
horizontal splitting: $h\mathfrak{g}=\mathfrak{so}(n)\oplus h\mathfrak{p,}$
when $[h\mathfrak{p},h\mathfrak{p}]\subset \mathfrak{so}(n)$ and $[\mathfrak{%
so}(n),h\mathfrak{p}]\subset h\mathfrak{p;}$ for the vertical splitting $v%
\mathfrak{g}=\mathfrak{so}(m)\oplus v\mathfrak{p,}$ when $[v\mathfrak{p},v%
\mathfrak{p}]\subset \mathfrak{so}(m)$ and $[\mathfrak{so}(m),v\mathfrak{p}%
]\subset v\mathfrak{p,}$ the Cartan d--connection determines an
N--anholonomic Riemannian structure on the nonholonomic bundle $\ ^{\alpha }%
\mathbf{\mathring{E}}=[hG=SO(n+1),$ $vG=SO(m+1),\;N_{i}^{e}].$ It is
possible to consider a quotient space with distinguished structure group $\
^{\alpha }\mathbf{V}_{\mathbf{N}}=\mathbf{G}/SO(n)\oplus $ $SO(m)$ regarding
$\mathbf{G}$ as a principal $\left( SO(n)\oplus SO(m)\right) $--bundle over $%
\ ^{\alpha }\mathbf{\mathring{E}},$ which is a N--anholonomic bundle. In
this case, we can always fix a local section of this bundle and pull--back $%
\;\;\ _{\mathbf{G}}^{\alpha }\widehat{\mathbf{\Gamma }}$ to give a $\left( h%
\mathfrak{g}\oplus v\mathfrak{g}\right) $--valued 1--form $\ _{\mathfrak{g}%
}^{\alpha }\widehat{\mathbf{\Gamma }}$ in a point $u\in \ ^{\alpha }\mathbf{%
\mathring{E}}.$ \footnote{%
There are involutive automorphisms $h\sigma =\pm 1$ and $v\sigma =\pm 1,$
respectively, of $h\mathfrak{g}$ and $v\mathfrak{g,}$ defined that $%
\mathfrak{so}(n)$ (or $\mathfrak{so}(m)$) is eigenspace $h\sigma =+1$ (or $%
v\sigma =+1)$ and $h\mathfrak{p}$ (or $v\mathfrak{p}$) is eigenspace $%
h\sigma =-1$ (or $v\sigma =-1).$ We construct a N--adapted fractional
decomposition taking into account the existing eigenspaces, when the
symmetric parts $\ ^{\alpha }\widehat{\mathbf{\Gamma }}\mathbf{\doteqdot }%
\frac{1}{2}\left( \ _{\mathfrak{g}}^{\alpha }\widehat{\mathbf{\Gamma }}%
\mathbf{+}\sigma \left( \ _{\mathfrak{g}}^{\alpha }\widehat{\mathbf{\Gamma }}%
\right) \right) ,$ with respective h- and v--splitting $\ ^{\alpha }\widehat{%
\mathbf{L}}\mathbf{\doteqdot }\frac{1}{2}\left( \ _{h\mathfrak{g}}^{\alpha }%
\widehat{\mathbf{L}}\mathbf{+}h\sigma \left( \ _{h\mathfrak{g}}^{\alpha }%
\widehat{\mathbf{L}}\right) \right)$ and $\ ^{\alpha }\widehat{\mathbf{C}}%
\mathbf{\doteqdot }\frac{1}{2}(\ _{v\mathfrak{g}}^{\alpha }\widehat{\mathbf{C%
}}+h\sigma (\ _{v\mathfrak{g}}^{\alpha }\widehat{\mathbf{C}})).$ This
defines a $\left( \mathfrak{so}(n)\oplus \mathfrak{so}(m)\right) $--valued
d--connection fractional 1--form. The antisymmetric part $\mathbf{e\doteqdot
}\frac{1}{2}\left( ^{\mathfrak{g}}\mathbf{\Gamma -}\sigma \left( ^{\mathfrak{%
g}}\mathbf{\Gamma }\right) \right) ,$ with respective h- and v--splitting $%
h\ ^{\alpha }\mathbf{e\doteqdot }\frac{1}{2}\left( \ _{h\mathfrak{g}%
}^{\alpha }\mathbf{e-}h\sigma \left( \ _{h\mathfrak{g}}^{\alpha }\mathbf{e}%
\right) \right) $ and $v\ ^{\alpha }\mathbf{e\doteqdot }\frac{1}{2}(\ _{v%
\mathfrak{g}}^{\alpha }\mathbf{e}-h\sigma (\ _{v\mathfrak{g}}^{\alpha }%
\mathbf{e})),$ defines a $\left( h\mathfrak{p}\oplus v\mathfrak{p}\right) $%
--valued N--adapted coframe for the Cartan--Killing inner product $<\cdot
,\cdot >_{\mathfrak{p}}$ on $\underline{T}_{u}\mathbf{G}\simeq h\mathfrak{g}%
\oplus v\mathfrak{g}$ restricted to $\underline{T}_{u}\mathbf{V}_{\mathbf{N}%
}\simeq \mathfrak{p.}$ This inner product, distinguished into h- and
v--components, provides a d--metric structure of type $\ ^{\alpha }\mathbf{g}%
=[\ ^{\alpha }g,\ ^{\alpha }h]$ (\ref{m1}),where $\ ^{\alpha }g=<h\ ^{\alpha
}\mathbf{e\otimes }h\ ^{\alpha }\mathbf{e}>_{h\mathfrak{p}}$ and $\ ^{\alpha
}h=<v\ ^{\alpha }\mathbf{e\otimes }v\ ^{\alpha }\mathbf{e}>_{v\mathfrak{p}}$
on $\ ^{\alpha }\mathbf{V}_{\mathbf{N}}=\mathbf{G}/SO(n)\oplus $ $SO(m).$}

It is possible to generate a $\mathbf{G(}=h\mathbf{G}\oplus v\mathbf{G)}$%
--invariant fractional d--derivative $\ ^{\alpha }\mathbf{D}$ with
restriction to the tangent space $\underline{T}\ ^{\alpha }\mathbf{V}_{%
\mathbf{N}}$ for any N--anholonomic curve flow $\gamma (\tau ,\mathbf{l})$
in $\ ^{\alpha }\mathbf{V}_{\mathbf{N}}=\mathbf{G}/SO(n)\oplus $ $SO(m)$ is
\begin{equation*}
\ ^{\alpha }\mathbf{D}_{\mathbf{X}}\ ^{\alpha }\mathbf{e=}\left[ \ ^{\alpha }%
\mathbf{e},\gamma _{\mathbf{l}}\rfloor \ ^{\alpha }\mathbf{\Gamma }\right] %
\mbox{\ and \ }\ ^{\alpha }\mathbf{D}_{\mathbf{Y}}\ ^{\alpha }\mathbf{e=}%
\left[ \ ^{\alpha }\mathbf{e},\gamma _{\mathbf{\tau }}\rfloor \ ^{\alpha }%
\mathbf{\Gamma }\right] ,
\end{equation*}%
admitting further h- and v--decompositions. The derivatives $\ ^{\alpha }%
\mathbf{D}_{\mathbf{X}}$ and $\ ^{\alpha }\mathbf{D}_{\mathbf{Y}}$ are
equivalent to (\ref{part01}) and obey the Cartan structure equations (\ref%
{mtors}) and (\ref{mcurv}).

We consider a N--adapted orthonormalized coframe $\ ^{\alpha }\mathbf{e}%
^{\alpha ^{\prime }}=(\ ^{\alpha }e^{i^{\prime }},\ ^{\alpha }\mathbf{e}%
^{a^{\prime }})$ identified with the $\left( h\mathfrak{p}\oplus v\mathfrak{p%
}\right) $--valued coframe $\ ^{\alpha }\mathbf{e}$ in a fixed orthonormal
basis for $\mathfrak{p=}h\mathfrak{p}\oplus v\mathfrak{p\subset }h\mathfrak{g%
}\oplus v\mathfrak{g.}$ For the kernel/ cokernel of Lie algebra
multiplications in the h- and v--subspaces, respectively, $\left[ \ ^{\alpha
}\mathbf{e}_{h\mathbf{X}},\cdot \right] _{h\mathfrak{g}}$ and $\left[ \
^{\alpha }\mathbf{e}_{v\mathbf{X}},\cdot \right] _{v\mathfrak{g}},$ we can
decompose the coframes into parallel and perpendicular parts with respect to
$\ ^{\alpha }\mathbf{e}_{\mathbf{X}},$ $\ ^{\alpha }\mathbf{e=(e}_{C}=h%
\mathbf{e}_{C}+v\mathbf{e}_{C},\mathbf{e}_{C^{\perp }}=h\mathbf{e}_{C^{\perp
}}+v\mathbf{e}_{C^{\perp }}\mathbf{),}$ for $\mathfrak{p(}=h\mathfrak{p}%
\oplus v\mathfrak{p)}$--valued mutually orthogonal d--vectors $\mathbf{e}%
_{C} $ \ and $\mathbf{e}_{C^{\perp }},$ when there are satisfied the
conditions $\left[ \mathbf{e}_{\mathbf{X}},\mathbf{e}_{C}\right] _{\mathfrak{%
g}}=0$ but $\left[ \mathbf{e}_{\mathbf{X}},\mathbf{e}_{C^{\perp }}\right] _{%
\mathfrak{g}}\neq 0;$ such conditions can be stated in h- and v--component
form, respectively, $\left[ h\mathbf{e}_{\mathbf{X}},h\mathbf{e}_{C}\right]
_{h\mathfrak{g}}=0,$ $\left[ h\mathbf{e}_{\mathbf{X}},h\mathbf{e}_{C^{\perp
}}\right] _{h\mathfrak{g}}\neq 0$ and $\left[ v\mathbf{e}_{\mathbf{X}},v%
\mathbf{e}_{C}\right] _{v\mathfrak{g}}=0,$ $\left[ v\mathbf{e}_{\mathbf{X}},v%
\mathbf{e}_{C^{\perp }}\right] _{v\mathfrak{g}}\neq 0.$ There are
decompositions
\begin{eqnarray*}
T_{u}\mathbf{V}_{\mathbf{N}} &\simeq &\mathfrak{p=}h\mathfrak{p}\oplus v%
\mathfrak{p}=\mathfrak{g=}h\mathfrak{g}\oplus v\mathfrak{g}/\mathfrak{so}%
(n)\oplus \mathfrak{so}(m) \\
\mbox{ \ and \ }\mathfrak{p} &\mathfrak{=p}&_{C}\oplus \mathfrak{p}%
_{C^{\perp }}=\left( h\mathfrak{p}_{C}\oplus v\mathfrak{p}_{C}\right) \oplus
\left( h\mathfrak{p}_{C^{\perp }}\oplus v\mathfrak{p}_{C^{\perp }}\right) ,
\end{eqnarray*}%
with $\mathfrak{p}_{\parallel }\subseteq \mathfrak{p}_{C}$ and $\mathfrak{p}%
_{C^{\perp }}\subseteq \mathfrak{p}_{\perp },$ where $\left[ \mathfrak{p}%
_{\parallel },\mathfrak{p}_{C}\right] =0,$ $<\mathfrak{p}_{C^{\perp }},%
\mathfrak{p}_{C}>=0,$ but $\left[ \mathfrak{p}_{\parallel },\mathfrak{p}%
_{C^{\perp }}\right] \neq 0$ (i.e. $\mathfrak{p}_{C}$ is the centralizer of $%
\mathbf{e}_{\mathbf{X}}$ in $\mathfrak{p=}h\mathfrak{p}\oplus v\mathfrak{%
p\subset }h\mathfrak{g}\oplus v\mathfrak{g);}$ in h- \ and v--components,
one have $h\mathfrak{p}_{\parallel }\subseteq h\mathfrak{p}_{C}$ and $h%
\mathfrak{p}_{C^{\perp }}\subseteq h\mathfrak{p}_{\perp },$ where $\left[ h%
\mathfrak{p}_{\parallel },h\mathfrak{p}_{C}\right] =0,$ $<h\mathfrak{p}%
_{C^{\perp }},h\mathfrak{p}_{C}>=0,$ but $\left[ h\mathfrak{p}_{\parallel },h%
\mathfrak{p}_{C^{\perp }}\right] \neq 0$ (i.e. $h\mathfrak{p}_{C}$ is the
centralizer of $\mathbf{e}_{h\mathbf{X}}$ in $h\mathfrak{p\subset }h%
\mathfrak{g)}$ and $v\mathfrak{p}_{\parallel }\subseteq v\mathfrak{p}_{C}$
and $v\mathfrak{p}_{C^{\perp }}\subseteq v\mathfrak{p}_{\perp },$ where $%
\left[ v\mathfrak{p}_{\parallel },v\mathfrak{p}_{C}\right] =0,$ $<v\mathfrak{%
p}_{C^{\perp }},v\mathfrak{p}_{C}>=0,$ but $\left[ v\mathfrak{p}_{\parallel
},v\mathfrak{p}_{C^{\perp }}\right] \neq 0$ (i.e. $v\mathfrak{p}_{C}$ is the
centralizer of $\mathbf{e}_{v\mathbf{X}}$ in $v\mathfrak{p\subset }v%
\mathfrak{g).}$

Acting with the canonical d--connection derivative $\ ^{\alpha }\widehat{%
\mathbf{D}}_{\mathbf{X}}$ of a d--covector perpendicular (or parallel) to $\
^{\alpha }\mathbf{e}_{\mathbf{X}},$ we get a new d--vector which is parallel
(or perpendicular) to $\ ^{\alpha }\mathbf{e}_{\mathbf{X}},$ i.e. $\
^{\alpha }\widehat{\mathbf{D}}\mathbf{e}_{C}\in \mathfrak{p}_{C^{\perp }}$
(or $\ ^{\alpha }\widehat{\mathbf{D}}\mathbf{e}_{C^{\perp }}\in \mathfrak{p}%
_{C}).$ In h- \ and v--components, such formulas are written $\ ^{\alpha }%
\widehat{\mathbf{D}}_{h\mathbf{X}}h\mathbf{e}_{C}\in h\mathfrak{p}_{C^{\perp
}}$ (or $\ ^{\alpha }\widehat{\mathbf{D}}_{h\mathbf{X}}h\mathbf{e}_{C^{\perp
}}\in h\mathfrak{p}_{C})$ and $\ ^{\alpha }\widehat{\mathbf{D}}_{v\mathbf{X}%
}v\mathbf{e}_{C}\in v\mathfrak{p}_{C^{\perp }}$ (or $\ ^{\alpha }\widehat{%
\mathbf{D}}_{v\mathbf{X}}v\mathbf{e}_{C^{\perp }}\in v\mathfrak{p}_{C}).$
All such d--algebraic relations can be written in N--anholonomic manifolds
and canonical d--connection settings, for instance, using certain relations
of type
\begin{equation*}
\ ^{\alpha }\widehat{\mathbf{D}}_{\mathbf{X}}(\mathbf{e}^{\alpha ^{\prime
}})_{C}=\mathbf{v}_{~\beta ^{\prime }}^{\alpha ^{\prime }}(\mathbf{e}^{\beta
^{\prime }})_{C^{\perp }}\mbox{ \ and \ }\ ^{\alpha }\widehat{\mathbf{D}}_{%
\mathbf{X}}(\mathbf{e}^{\alpha ^{\prime }})_{C^{\perp }}=-\mathbf{v}_{~\beta
^{\prime }}^{\alpha ^{\prime }}(\mathbf{e}^{\beta ^{\prime }})_{C},
\end{equation*}%
for some antisymmetric d--tensors $\mathbf{v}^{\alpha ^{\prime }\beta
^{\prime }}=-\mathbf{v}^{\beta ^{\prime }\alpha ^{\prime }}.$ This is a
N--adapted $\left( SO(n)\oplus SO(m)\right) $--parallel frame defining a
generalization of the concept of parallel frame on N--adapted fractional
manifolds whenever $\mathfrak{p}_{C}$ is larger than $\mathfrak{p}%
_{\parallel }.$ If we substitute $\ ^{\alpha }\mathbf{e}^{\alpha ^{\prime
}}=(\ ^{\alpha }e^{i^{\prime }},\ ^{\alpha }\mathbf{e}^{a^{\prime }})$ into
the last formulas and considering h- and v--components, we construct $SO(n)$%
--parallel and $SO(m)$--parallel frames.

\setcounter{equation}{0} \renewcommand{\theequation}
{A.\arabic{equation}} \setcounter{subsection}{0}
\renewcommand{\thesubsection}
{A.\arabic{subsection}}
\section{Conclusions}

The use of the fractional calculus techniques in differential geometry is still at the beginning of its application.
The fractional operators reveal a complex structure and possess less properties that the classical ones.
From these reasons the fractional operators are not easy  to be used within the differential geometry and its applications.
In this manuscript the calculations were done within  Caputo derivative.
It was shown that for corresponding classes of  nonholonomic distributions a large class of physical theories are modelled as nonholonomic manifolds possessing constant matrix curvature. As a result  we encoded the fractional dynamics of interactions and constraints into the geometry of curve flows and solitonic hierarchies.

\appendix

\section{Fractional Caputo N--anholonomic Manifolds}

\label{sappa}For spaces of fractional dimension, it is possible to construct
models of fractional differential geometry similarly to certain
corresponding integer dimension geometries if the Caputo fractional
derivative is used \cite{vrfrf,vrfrg}. Applying nonholonomic deformations,
the constructions can be generalized for another types of fractional
derivatives.

\subsection{Caputo fractional derivatives, local (co) bases and integration}

The fractional left, respectively, right Caputo derivatives are defined by
formulas
\begin{eqnarray}
&&\ _{\ _{1}x}\overset{\alpha }{\underline{\partial }}_{x}f(x):=\frac{1}{%
\Gamma (s-\alpha )}\int\limits_{\ \ _{1}x}^{x}(x-\ x^{\prime })^{s-\alpha
-1}\left( \frac{\partial }{\partial x^{\prime }}\right) ^{s}f(x^{\prime
})dx^{\prime };  \label{lfcd} \\
&&\ _{\ x}\overset{\alpha }{\underline{\partial }}_{\ _{2}x}f(x):=\frac{1}{%
\Gamma (s-\alpha )}\int\limits_{x}^{\ _{2}x}(x^{\prime }-x)^{s-\alpha
-1}\left( -\frac{\partial }{\partial x^{\prime }}\right) ^{s}f(x^{\prime
})dx^{\prime }\ .  \notag
\end{eqnarray}%
We can introduce $\ \overset{\alpha }{d}:=(dx^{j})^{\alpha }\ \ _{\ 0}%
\overset{\alpha }{\underline{\partial }}_{j}$ for the fractional absolute
differential, where $\ \overset{\alpha }{d}x^{j}=(dx^{j})^{\alpha }\frac{%
(x^{j})^{1-\alpha }}{\Gamma (2-\alpha )}$ if $\ _{1}x^{i}=0.$ Such formulas
allow us to elaborate the concept of fractional tangent bundle $\overset{%
\alpha }{\underline{T}}M,$ \ for $\alpha \in (0,1),$ associated to a
manifold $M$ of necessary smooth class and integer $\dim M=n.$\footnote{%
For simplicity, we may write both the integer and fractional local
coordinates in the form $u^{\beta }=(x^{j},y^{a}).$ We underlined the symbol
$T$ in order to emphasize that we shall associate the approach to a
fractional Caputo derivative.} \

Let us denote by $L_{z}(\ _{1}x,\ _{2}x)$ the set of those Lesbegue
measurable functions $f$ on $[\ _{1}x,\ _{2}x]$ \ when $||f||_{z}=(\int%
\limits_{_{1}x}^{_{2}x}|f(x)|^{z}dx)^{1/z}<\infty $ and $C^{z}[\ _{1}x,\
_{2}x]$ be the space of functions which are $z$ times continuously
differentiable on this interval. For any real--valued function $f(x)$
defined on a closed interval $[\ _{1}x,\ _{2}x],$ there is a function $%
F(x)=_{\ _{1}x}\overset{\alpha }{I}_{x}\ f(x)$ defined by the fractional
Riemann--Liouville integral $\ _{\ _{1}x}\overset{\alpha }{I}_{x}f(x):=\frac{%
1}{\Gamma (\alpha )}\int\limits_{_{1}x}^{x}(x-x^{\prime })^{\alpha
-1}f(x^{\prime })dx^{\prime },$ when $f(x)=\ _{\ _{1}x}\overset{\alpha }{%
\underline{\partial }}_{x}F(x),$ for all $x\in \lbrack \ _{1}x,\ _{2}x],$
satisfies the conditions
\begin{eqnarray*}\label{aux01}
\ _{\ _{1}x}\overset{\alpha }{\underline{\partial }}_{x}\left( _{\ _{1}x}%
\overset{\alpha }{I}_{x}f(x)\right) &=&f(x),\ \alpha >0, \\
_{\ _{1}x}\overset{\alpha }{I}_{x}\left( \ _{\ _{1}x}\overset{\alpha }{%
\underline{\partial }}_{x}F(x)\right) &=&F(x)-F(\ _{1}x),\ 0<\alpha <1.
\end{eqnarray*}

We can consider fractional (co) frame bases on $\overset{\alpha }{\underline{%
T}}M.$ For instance, a fractional frame basis $\overset{\alpha }{\underline{e%
}}_{\beta }=e_{\ \beta }^{\beta ^{\prime }}(u^{\beta })\overset{\alpha }{%
\underline{\partial }}_{\beta ^{\prime }}$\ is connected via a vierlbein
transform $e_{\ \beta }^{\beta ^{\prime }}(u^{\beta })$ with a fractional
local coordinate basis
\begin{equation}
\overset{\alpha }{\underline{\partial }}_{\beta ^{\prime }}=\left( \overset{%
\alpha }{\underline{\partial }}_{j^{\prime }}=_{\ _{1}x^{j^{\prime }}}%
\overset{\alpha }{\underline{\partial }}_{j^{\prime }},\overset{\alpha }{%
\underline{\partial }}_{b^{\prime }}=_{\ _{1}y^{b^{\prime }}}\overset{\alpha
}{\underline{\partial }}_{b^{\prime }}\right) ,  \label{frlcb}
\end{equation}%
for $j^{\prime }=1,2,...,n$ and $b^{\prime }=n+1,n+2,...,n+n.$ The
fractional co--bases are related via $\overset{\alpha }{\underline{e}}^{\
\beta }=e_{\beta ^{\prime }\ }^{\ \beta }(u^{\beta })\overset{\alpha }{d}%
u^{\beta ^{\prime }},$ where
\begin{equation}
\ _{\ }\overset{\alpha }{d}u^{\beta ^{\prime }}=\left( (dx^{i^{\prime
}})^{\alpha },(dy^{a^{\prime }})^{\alpha }\right) .  \label{frlccb}
\end{equation}

The fractional absolute differential $\overset{\alpha }{d}$ is written in
the form
\begin{equation*}
\overset{\alpha }{d}:=(dx^{j})^{\alpha }\ \ _{\ 0}\overset{\alpha }{%
\underline{\partial }}_{j},\mbox{ where }\ \overset{\alpha }{d}%
x^{j}=(dx^{j})^{\alpha }\frac{(x^{j})^{1-\alpha }}{\Gamma (2-\alpha )},
\end{equation*}%
where we consider $\ _{1}x^{i}=0.$ The differentials $dx^{j}=(dx^{j})^{%
\alpha =1}$ are used as local coordinate co-bases/--frames or the
''integer'' calculus. For $0<\alpha <1,$ we have $dx=(dx)^{1-\alpha
}(dx)^{\alpha }.$ $\ $The ''fractional'' symbol $(dx^{j})^{\alpha }$ is
related to $\overset{\alpha }{d}x^{j}$ and can be used instead of
''integer'' $dx^{i}$ for elaborating a co--vector/differential form
calculus. Following the above system of notation, the exterior fractional
differential is
\begin{equation*}
\overset{\alpha }{d}=\sum\limits_{j=1}^{n}\Gamma (2-\alpha )(x^{j})^{\alpha
-1}\ \overset{\alpha }{d}x^{j}\ \ _{\ 0}\overset{\alpha }{\underline{%
\partial }}_{j}.
\end{equation*}

The fractional integration for differential forms on interval $L=[\ _{1}x,\
_{2}x]$ is performed following formula
\begin{equation*}
\ _{L}\overset{\alpha }{I}[x]\ \ _{\ _{1}x}\overset{\alpha }{d}_{x}f(x)=f(\
_{2}x)-f(\ _{1}x),
\end{equation*}%
when the fractional differential of a function $\ f(x)$ is $_{\ _{1}x}%
\overset{\alpha }{d}_{x}f(x)=[...],$ when
\begin{equation*}
\int\limits_{_{1}x}^{_{2}x}\frac{(dx)^{1-\alpha }}{\Gamma (\alpha )(\
_{2}x-x)^{1-\alpha }}[(dx^{\prime })^{\alpha }\ _{\ _{1}x}\overset{\alpha }{%
\underline{\partial }}_{x^{\prime \prime }}f(x^{\prime \prime })]=f(x)-f(\
_{1}x).
\end{equation*}

The nonholonomic geometry of fractional tangent bundle depends on the type
of chosen fractional derivative. We also emphasize that the above formulas
can be generalized for an arbitrary vector bundle $E$ and/or nonholonomic
manifold $\mathbf{V}.$

\subsection{N-- and d--connections and metrics}

A nonlinear connection (N--connection) $\overset{\alpha }{\mathbf{N}}$ \ for
a fractional space $\overset{\alpha }{\mathbf{V}}$ is defined by a
nonholonomic distribution (Whitney sum) with conventional h-- and
v--subspaces, $\underline{h}\overset{\alpha }{\mathbf{V}}$ and $\underline{v}%
\overset{\alpha }{\mathbf{V}},$%
\begin{equation}
\overset{\alpha }{\underline{T}}\overset{\alpha }{\mathbf{V}}=\underline{h}%
\overset{\alpha }{\mathbf{V}}\mathbf{\oplus }\underline{v}\overset{\alpha }{%
\mathbf{V}}.  \label{whit}
\end{equation}

A fractional N--connection is defined by its local coefficients $\overset{%
\alpha }{\mathbf{N}}\mathbf{=}\{\ ^{\alpha }N_{i}^{a}\},$ when
\begin{equation*}
\overset{\alpha }{\mathbf{N}}\mathbf{=}\ ^{\alpha
}N_{i}^{a}(u)(dx^{i})^{\alpha }\otimes \overset{\alpha }{\underline{\partial
}}_{a}.
\end{equation*}%
For a N--connection $\overset{\alpha }{\mathbf{N}},$ we can always construct
a class of fractional (co) frames (N--adapted) linearly depending on $\
^{\alpha }N_{i}^{a},$
\begin{eqnarray}
\ ^{\alpha }\mathbf{e}_{\beta } &=&\left[ \ ^{\alpha }\mathbf{e}_{j}=\overset%
{\alpha }{\underline{\partial }}_{j}-\ ^{\alpha }N_{j}^{a}\overset{\alpha }{%
\underline{\partial }}_{a},\ ^{\alpha }e_{b}=\overset{\alpha }{\underline{%
\partial }}_{b}\right] ,  \label{dder} \\
\ ^{\alpha }\mathbf{e}^{\beta } &=&[\ ^{\alpha }e^{j}=(dx^{j})^{\alpha },\
^{\alpha }\mathbf{e}^{b}=(dy^{b})^{\alpha }+\ ^{\alpha
}N_{k}^{b}(dx^{k})^{\alpha }].  \label{ddif}
\end{eqnarray}%
The nontrivial nonholonomy coefficients are computed $\ ^{\alpha }W_{ib}^{a}=%
\overset{\alpha }{\underline{\partial }}_{b}\ ^{\alpha }N_{i}^{a}$ and $\
^{\alpha }W_{ij}^{a}=\ ^{\alpha }\Omega _{ji}^{a}=\ ^{\alpha }\mathbf{e}%
_{i}\ ^{\alpha }N_{j}^{a}-\ ^{\alpha }\mathbf{e}_{j}\ ^{\alpha }N_{i}^{a}$
for
\begin{equation*}
\left[ \ ^{\alpha }\mathbf{e}_{\alpha },\ ^{\alpha }\mathbf{e}_{\beta }%
\right] =\ ^{\alpha }\mathbf{e}_{\alpha }\ ^{\alpha }\mathbf{e}_{\beta }-\
^{\alpha }\mathbf{e}_{\beta }\ ^{\alpha }\mathbf{e}_{\alpha }=\ ^{\alpha
}W_{\alpha \beta }^{\gamma }\ ^{\alpha }\mathbf{e}_{\gamma }.
\end{equation*}%
In above formulas, the values $\ ^{\alpha }\Omega _{ji}^{a}$ are called the
coefficients of N--connection curvature. A nonholonomic manifold defined by
a structure $\overset{\alpha }{\mathbf{N}}$ is called, in brief, a
N--anholonomic fractional manifold.

We introduce a metric structure $\ \overset{\alpha }{\mathbf{g}}=\{\
^{\alpha }g_{\underline{\alpha }\underline{\beta }}\}$ on $\overset{\alpha }{%
\mathbf{V}}$ \ as a symmetric second rank tensor with coefficients
determined locally with respect to a corresponding tensor product of
fractional differentials,
\begin{equation}
\overset{\alpha }{\mathbf{g}}=\ ^{\alpha }g_{\underline{\gamma }\underline{%
\beta }}(u)(du^{\underline{\gamma }})^{\alpha }\otimes (du^{\underline{\beta
}})^{\alpha }.  \label{fmcf}
\end{equation}%
For N--adapted constructions, it is important to use the property that any
fractional metric $\overset{\alpha }{\mathbf{g}}$ can be represented
equivalently as a distinguished metric (d--metric), $\ \overset{\alpha }{%
\mathbf{g}}=\left[ \ ^{\alpha }g_{kj},\ ^{\alpha }g_{cb}\right] ,$ when
\begin{eqnarray}
\ \overset{\alpha }{\mathbf{g}} &=&\ ^{\alpha }g_{kj}(x,y)\ ^{\alpha
}e^{k}\otimes \ ^{\alpha }e^{j}+\ ^{\alpha }g_{cb}(x,y)\ ^{\alpha }\mathbf{e}%
^{c}\otimes \ ^{\alpha }\mathbf{e}^{b}  \label{m1} \\
&=&\eta _{k^{\prime }j^{\prime }}\ ^{\alpha }e^{k^{\prime }}\otimes \
^{\alpha }e^{j^{\prime }}+\eta _{c^{\prime }b^{\prime }}\ ^{\alpha }\mathbf{e%
}^{c^{\prime }}\otimes \ ^{\alpha }\mathbf{e}^{b^{\prime }},  \notag
\end{eqnarray}%
where matrices $\eta _{k^{\prime }j^{\prime }}=diag[\pm 1,\pm 1,...,\pm 1]$
and $\eta _{a^{\prime }b^{\prime }}=diag[\pm 1,\pm 1,...,\pm 1],$ for the
signature of a ''prime'' spacetime $\mathbf{V,}$ are obtained by frame
transforms $\eta _{k^{\prime }j^{\prime }}=e_{\ k^{\prime }}^{k}\ e_{\
j^{\prime }}^{j}\ _{\ }^{\alpha }g_{kj}$ and $\eta _{a^{\prime }b^{\prime
}}=e_{\ a^{\prime }}^{a}\ e_{\ b^{\prime }}^{b}\ _{\ }^{\alpha }g_{ab}.$

A distinguished connection (d--connection) $\overset{\alpha }{\mathbf{D}}$
on $\overset{\alpha }{\mathbf{V}}$ is defined as a linear connection
preserving under parallel transports the Whitney sum (\ref{whit}). We can
associate a N--adapted differential 1--form
\begin{equation}
\ ^{\alpha }\mathbf{\Gamma }_{\ \beta }^{\tau }=\ ^{\alpha }\mathbf{\Gamma }%
_{\ \beta \gamma }^{\tau }\ ^{\alpha }\mathbf{e}^{\gamma },  \label{fdcf}
\end{equation}%
parametrizing the coefficients (with respect to (\ref{ddif}) and (\ref{dder}%
)) in the form $\ ^{\alpha }\mathbf{\Gamma }_{\ \tau \beta }^{\gamma
}=\left( \ ^{\alpha }L_{jk}^{i},\ ^{\alpha }L_{bk}^{a},\ ^{\alpha
}C_{jc}^{i},\ ^{\alpha }C_{bc}^{a}\right) .$

The absolute fractional differential $\ ^{\alpha }\mathbf{d}=\ _{\ _{1}x}%
\overset{\alpha }{d}_{x}+\ _{\ _{1}y}\overset{\alpha }{d}_{y}$ acts on
fractional differential forms in N--adapted form; the value $\ ^{\alpha }%
\mathbf{d:=}\ ^{\alpha }\mathbf{e}^{\beta }\ ^{\alpha }\mathbf{e}_{\beta }$
splits into exterior h- and v--derivatives when
\begin{equation*}
\ _{\ _{1}x}\overset{\alpha }{d}_{x}:=(dx^{i})^{\alpha }\ \ _{\ _{1}x}%
\overset{\alpha }{\underline{\partial }}_{i}=\ ^{\alpha }e^{j}\ ^{\alpha }%
\mathbf{e}_{j}\mbox{ and }_{\ _{1}y}\overset{\alpha }{d}_{y}:=(dy^{a})^{%
\alpha }\ \ _{\ _{1}x}\overset{\alpha }{\underline{\partial }}_{a}=\
^{\alpha }\mathbf{e}^{b}\ ^{\alpha }e_{b}.
\end{equation*}

The torsion and curvature of a fractional d--connection $\overset{\alpha }{%
\mathbf{D}}=\{\ ^{\alpha }\mathbf{\Gamma }_{\ \beta \gamma }^{\tau }\}$ can
be defined and computed, respectively, as fractional 2--forms,
\begin{eqnarray}
\ ^{\alpha }\mathcal{T}^{\tau } &\doteqdot &\overset{\alpha }{\mathbf{D}}\
^{\alpha }\mathbf{e}^{\tau }=\ ^{\alpha }\mathbf{d}\ ^{\alpha }\mathbf{e}%
^{\tau }+\ ^{\alpha }\mathbf{\Gamma }_{\ \beta }^{\tau }\wedge \ ^{\alpha }%
\mathbf{e}^{\beta }\mbox{ and }  \label{tors} \\
\ ^{\alpha }\mathcal{R}_{~\beta }^{\tau } &\doteqdot &\overset{\alpha }{%
\mathbf{D}}\mathbf{\ ^{\alpha }\Gamma }_{\ \beta }^{\tau }=\ ^{\alpha }%
\mathbf{d\ ^{\alpha }\Gamma }_{\ \beta }^{\tau }-\ ^{\alpha }\mathbf{\Gamma }%
_{\ \beta }^{\gamma }\wedge \ ^{\alpha }\mathbf{\Gamma }_{\ \gamma }^{\tau
}=\ ^{\alpha }\mathbf{R}_{\ \beta \gamma \delta }^{\tau }\ ^{\alpha }\mathbf{%
e}^{\gamma }\wedge \ ^{\alpha }\mathbf{e}^{\delta }.  \notag
\end{eqnarray}

There are two another important geometric objects: the fractional Ricci
tensor $\ ^{\alpha }\mathcal{R}ic=\{\ ^{\alpha }\mathbf{R}_{\alpha \beta
}\doteqdot \ ^{\alpha }\mathbf{R}_{\ \alpha \beta \tau }^{\tau }\}$ with
components
\begin{equation}
\ ^{\alpha }R_{ij}\doteqdot \ ^{\alpha }R_{\ ijk}^{k},\ \ \ ^{\alpha
}R_{ia}\doteqdot -\ ^{\alpha }R_{\ ika}^{k},\ \ ^{\alpha }R_{ai}\doteqdot \
^{\alpha }R_{\ aib}^{b},\ \ ^{\alpha }R_{ab}\doteqdot \ ^{\alpha }R_{\
abc}^{c}  \label{dricci}
\end{equation}%
and the scalar curvature of fractional d--connection $\overset{\alpha }{%
\mathbf{D}},$
\begin{equation}
\ _{s}^{\alpha }\mathbf{R}\doteqdot \ ^{\alpha }\mathbf{g}^{\tau \beta }\
^{\alpha }\mathbf{R}_{\tau \beta }=\ ^{\alpha }R+\ ^{\alpha }S,\ ^{\alpha
}R=\ ^{\alpha }g^{ij}\ ^{\alpha }R_{ij},\ \ ^{\alpha }S=\ ^{\alpha }g^{ab}\
^{\alpha }R_{ab},  \label{sdccurv}
\end{equation}%
with $\ ^{\alpha }\mathbf{g}^{\tau \beta }$ being the inverse coefficients
to a d--metric (\ref{m1}).

We can introduce the Einstein tensor $\ ^{\alpha }\mathcal{E}ns=\{\ ^{\alpha
}\mathbf{G}_{\alpha \beta }\},$
\begin{equation}
\ ^{\alpha }\mathbf{G}_{\alpha \beta }:=\ ^{\alpha }\mathbf{R}_{\alpha \beta
}-\frac{1}{2}\ ^{\alpha }\mathbf{g}_{\alpha \beta }\ \ _{s}^{\alpha }\mathbf{%
R.}  \label{enstdt}
\end{equation}

For various applications, we can considered more special classes of
d--connections:

\begin{itemize}
\item There is a unique canonical metric compatible fractional d--connection
$\ ^{\alpha }\widehat{\mathbf{D}}=\{\ ^{\alpha }\widehat{\mathbf{\Gamma }}%
_{\ \alpha \beta }^{\gamma }=\left( \ ^{\alpha }\widehat{L}_{jk}^{i},\
^{\alpha }\widehat{L}_{bk}^{a},\ ^{\alpha }\widehat{C}_{jc}^{i},\ ^{\alpha }%
\widehat{C}_{bc}^{a}\right) \},$ when $\ ^{\alpha }\widehat{\mathbf{D}}\
\left( \ ^{\alpha }\mathbf{g}\right) =0,$ satisfying the conditions $\
^{\alpha }\widehat{T}_{\ jk}^{i}=0$ and $\ ^{\alpha }\widehat{T}_{\
bc}^{a}=0,$ but $\ ^{\alpha }\widehat{T}_{\ ja}^{i},\ ^{\alpha }\widehat{T}%
_{\ ji}^{a}$ and $\ ^{\alpha }\widehat{T}_{\ bi}^{a}$ are not zero. The
N--adapted coefficients are explicitly determined by the coefficients of (%
\ref{m1}),
\begin{eqnarray}
\ ^{\alpha }\widehat{L}_{jk}^{i} &=&\frac{1}{2}\ ^{\alpha }g^{ir}\left( \
^{\alpha }\mathbf{e}_{k}\ ^{\alpha }g_{jr}+\ ^{\alpha }\mathbf{e}_{j}\
^{\alpha }g_{kr}-\ ^{\alpha }\mathbf{e}_{r}\ ^{\alpha }g_{jk}\right) ,
\notag \\
\ ^{\alpha }\widehat{L}_{bk}^{a} &=&\ ^{\alpha }e_{b}(\ ^{\alpha }N_{k}^{a})+%
\frac{1}{2}\ ^{\alpha }g^{ac}(\ ^{\alpha }\mathbf{e}_{k}\ ^{\alpha }g_{bc}
\notag \\
&&-\ ^{\alpha }g_{dc}\ \ ^{\alpha }e_{b}\ ^{\alpha }N_{k}^{d}-\ ^{\alpha
}g_{db}\ \ ^{\alpha }e_{c}\ ^{\alpha }N_{k}^{d}),  \notag \\
\ ^{\alpha }\widehat{C}_{jc}^{i} &=&\frac{1}{2}\ ^{\alpha }g^{ik}\ ^{\alpha
}e_{c}\ ^{\alpha }g_{jk},  \label{candcon} \\
\ \ ^{\alpha }\widehat{C}_{bc}^{a} &=&\frac{1}{2}\ ^{\alpha }g^{ad}\left( \
^{\alpha }e_{c}\ ^{\alpha }g_{bd}+\ ^{\alpha }e_{c}\ ^{\alpha }g_{cd}-\
^{\alpha }e_{d}\ ^{\alpha }g_{bc}\right) .  \notag
\end{eqnarray}

\item The fractional Levi--Civita connection $\ ^{\alpha }\nabla =\{\ \
^{\alpha }\Gamma _{\ \alpha \beta }^{\gamma }\}$ can be defined in standard
from but for the fractional Caputo left derivatives acting on the
coefficients of a fractional metric (\ref{fmcf}).
\end{itemize}

On spaces with nontrivial nonholonomic structure, it is preferred to work on
$\overset{\alpha }{\mathbf{V}}$ with $\ ^{\alpha }\widehat{\mathbf{D}}=\{\
^{\alpha }\widehat{\mathbf{\Gamma }}_{\ \tau \beta }^{\gamma }\}$ instead of
$\ ^{\alpha }\nabla $ (the last one is not adapted to the N--connection
splitting (\ref{whit})). The torsion $\ ^{\alpha }\widehat{\mathcal{T}}%
^{\tau }$ (\ref{tors}) \ of $\ ^{\alpha }\widehat{\mathbf{D}}$ is uniquely
induced nonholonomically by off--diagonal coefficients of the d--metric (\ref%
{m1}).

With respect to N--adapted fractional bases (\ref{dder}) and (\ref{ddif}),
the coefficients of the fractional Levi--Civita and canonical d--connection
satisfy the distorting relations
\begin{equation}
\ ^{\alpha }\Gamma _{\ \alpha \beta }^{\gamma }=\ ^{\alpha }\widehat{\mathbf{%
\Gamma }}_{\ \alpha \beta }^{\gamma }+\ \ ^{\alpha }Z_{\ \alpha \beta
}^{\gamma },  \label{cdeft}
\end{equation}%
where the N--adapted coefficients of distortion tensor $\ Z_{\ \alpha \beta
}^{\gamma }$ \ are computed%
\begin{eqnarray*}
\ \ \ ^{\alpha }Z_{jk}^{i} &=&0,\ \ \ ^{\alpha }Z_{jk}^{a}=-\ \ ^{\alpha
}C_{jb}^{i}\ \ ^{\alpha }g_{ik}\ \ ^{\alpha }g^{ab}-\frac{1}{2}\ \ ^{\alpha
}\Omega _{jk}^{a}, \\
\ \ \ ^{\alpha }Z_{bk}^{i} &=&\frac{1}{2}\ \ ^{\alpha }\Omega _{jk}^{c}\ \
^{\alpha }g_{cb}\ \ ^{\alpha }g^{ji}-\frac{1}{2}(\delta _{j}^{i}\delta
_{k}^{h}-\ \ ^{\alpha }g_{jk}\ \ ^{\alpha }g^{ih})\ \ ^{\alpha }C_{hb}^{j},
\\
\ \ \ ^{\alpha }Z_{bk}^{a} &=&\frac{1}{2}(\delta _{c}^{a}\delta _{d}^{b}+\ \
^{\alpha }g_{cd}\ \ ^{\alpha }g^{ab})\left[ \ \ ^{\alpha }L_{bk}^{c}-\ \
^{\alpha }e_{b}(\ \ ^{\alpha }N_{k}^{c})\right] , \\
\ \ \ ^{\alpha }Z_{kb}^{i} &=&\frac{1}{2}\ \ ^{\alpha }\Omega _{jk}^{a}\ \
^{\alpha }g_{cb}\ \ ^{\alpha }g^{ji}+\frac{1}{2}(\delta _{j}^{i}\delta
_{k}^{h}-\ \ ^{\alpha }g_{jk}\ \ ^{\alpha }g^{ih})\ \ ^{\alpha }C_{hb}^{j},
\\
\ ^{\alpha }Z_{jb}^{a} &=&-\frac{1}{2}(\delta _{c}^{a}\delta _{b}^{d}-\ \
^{\alpha }g_{cb}\ \ ^{\alpha }g^{ad})\left[ \ ^{\alpha }L_{dj}^{c}-\ \
^{\alpha }e_{d}(\ \ ^{\alpha }N_{j}^{c})\right] ,\ ^{\alpha }Z_{bc}^{a}=0, \\
\ ^{\alpha }Z_{ab}^{i} &=&-\frac{\ \ ^{\alpha }g^{ij}}{2}\{\left[ \ ^{\alpha
}L_{aj}^{c}-\ ^{\alpha }e_{a}(\ ^{\alpha }N_{j}^{c})\right] \ ^{\alpha
}g_{cb} \\
&&+\left[ \ ^{\alpha }L_{bj}^{c}-\ ^{\alpha }e_{b}(\ ^{\alpha }N_{j}^{c})%
\right] \ ^{\alpha }g_{ca}\}.
\end{eqnarray*}

\end{document}